\documentclass[%
 reprint,
 amsmath,amssymb,
 aps,
pra,
]{revtex4-2}

\usepackage{graphicx}
\usepackage{dcolumn}
\usepackage{bm}
\usepackage{hyperref}
\usepackage[mathlines]{lineno}
\usepackage{cleveref}
\usepackage{booktabs}
\usepackage{mathtools}
\usepackage{relsize}
\usepackage{physics}
\usepackage{mathbbol}

 \usepackage{tikz}
 \usepackage{lipsum}

\begin{document}

\title{Optimal observables for (non-)equilibrium quantum \nobreak{metrology} from the master equation}

\author{V\'ictor L\'opez-Pardo}
\affiliation{Instituto Galego de Física de Altas Enerxías (IGFAE), Universidade de Santiago de Compostela, Santiago de Compostela, 15782, Galicia, Spain}
\email{victorlopez.pardo@usc.es}

\author{Alexander Rothkopf}
\affiliation{Department of Physics, Korea University, Seoul 02841, Republic of Korea}
\email{akrothkopf@korea.ac.kr}

\date{\today}

\begin{abstract}
We demonstrate how observables with optimal sensitivity to environmental properties can be constructed explicitly from the master equation of an open quantum system. Our approach does not rely on the explicit solution of the master equation and instead expresses the symmetric logarithmic derivative (SLD), the operator of optimal sensitivity and key quantity in quantum metrology, solely in terms of expectation values. This makes the SLD available to a large class of systems of interest, both in and out of equilibrium. We validate our approach by reproducing the SLD for temperature in quantum Brownian motion and demonstrate its versatility by constructing the optimal observable for the non-equilibrium relaxation rate.
\end{abstract}


\maketitle


\section{Introduction}

Quantum metrology exploits the quantum properties of systems to estimate physical parameters with the highest possible precision, potentially surpassing classical limits (for a recent review, see Ref.~\cite{montenegro_review_2025}). A central object in local estimation theory \cite{paris_quantum_2009} is the \emph{symmetric logarithmic derivative} (SLD), \(\hat\Lambda_\theta\) \cite{helstrom_quantum_1969,holevo_probabilistic_2011}, which determines the \emph{quantum Fisher information} (QFI), a measure of the ultimate sensitivity with which a parameter \(\theta\) encoded in a quantum state \(\hat\rho(\theta)\) can be estimated. Defined implicitly via the {\it generalized Lyapunov equation}
\begin{equation}
\partial_\theta \hat \rho(\theta) = \frac{1}{2} \left( \hat\Lambda_\theta \hat\rho(\theta) + \hat\rho(\theta) \hat\Lambda_\theta \right) - \hat\rho(\theta)\langle \hat \Lambda_\theta \rangle,\label{eq:SLDdef}
\end{equation}
the SLD plays a dual role: it determines both the QFI via its variance \(\mathcal{F}_\theta = {\rm Var}[\hat\Lambda_\theta]\) , and more fundamentally, it constitutes the \emph{optimal observable} whose variance bounds the achievable estimation error in the quantum Cramér-Rao inequality \cite{braunstein_statistical_1994}.

Despite its central conceptual role, the practical use of the SLD in quantum metrology remains constrained \cite{montenegro_review_2025}. Its computation is often contingent on prior knowledge of the \emph{explicit solution} \(\hat\rho(\theta)\) to the dynamical equation governing the quantum system, readily available predominantly in Gaussian states \cite{Cenni:2021cfk} or two-level systems. This limitation is particularly restrictive in \emph{open quantum systems} \cite{breuer2002theory}, used for impurity metrology, where the explicit density matrix is typically available only through numerical simulations or perturbative expansions, such as in the study of heavy quark impurities in high-energy nuclear physics \cite{Akamatsu:2011se,Brambilla:2016wgg,Miura:2019ssi,Alund:2020ctu}. 

In the context of \emph{equilibrium thermometry}, for instance, it is common to evaluate the QFI for thermal (Gibbs) states by exploiting analytic forms of \(\hat\rho(T)\). Formally the SLD may be expressed as a simple function of the system Hamiltonian, such as \(\hat\Lambda_T = (\hat H - \langle H \rangle)/T^2\) \cite{potts_fundamental_2019} but direct energy measurements may not be feasible in practice. An interesting example is Ref.~\cite{mehboudi_using_2019} where the SLD for a heavy bosonic impurity was explicitly constructed 
\begin{align}
 \hat{\Lambda}_T &= C_x \left( \hat{x}^2 - \langle {x}^2 \rangle \right) + C_p \left( \hat{p}^2 - \langle {p}^2 \rangle \right)   
\end{align}
making reference to the expectation values of practically accessible operators, such as position $\hat x$ and momentum $\hat p$ in the coefficients
\begin{align}
    C_x &= \frac{4 \langle {p}^2 \rangle^2 \, \chi_T(\hat{x}^2) + \hbar^2 \chi_T(\hat{p}^2)}{8 \langle {x}^2 \rangle^2 \langle {p}^2 \rangle^2 - \hbar^4/2},
\end{align}
and $C_p$ for which one replaces $\hat x \leftrightarrow \hat p$ in $C_x$.
Note however that in the above expression the temperature susceptibilities $\chi_T(\hat{x}^2)$ and $\chi_T(\hat{p}^2)$ enter, which require a microscopic model and access to the explicit density matrix to produce predictions for $\chi_\theta (\hat {\cal O}) = \partial_\theta {\rm Tr}[\hat\rho(\theta)\hat{\cal O}]$.

In \emph{non-equilibrium thermometry}, the SLD has been investigated for various two-level (qubit) systems. Let us consider the interesting example of {\it in situ} thermometry with dephasing impurities pioneered in Ref.~\cite{Mitchison:2020szc}. In this scenario, due to exact energy conservation of the probe, thermalization with the surrounding environment is suppressed. Even in the absence of thermalization the probe can develop correlations with the environment that give access to temperature and an SLD has been constructed
\begin{align}
\hat{\Lambda}_T &\propto \cos(\varphi)\, \hat{\sigma}_\parallel + \sin(\varphi)\, \hat{\sigma}_\perp, \\
\tan(\varphi) &= \frac{|v|(1 - |v|)^2 \, \partial_T \phi}{\partial_T |v|}.
\end{align}
Here $|v|$ and $\phi$ parameterize the density matrix of the probe and $\hat{\sigma}_\parallel$ and $\hat{\sigma}_\perp$ denote linear combinations of spin measurement operators in the $x$ and $y$ direction. Again we see that derivatives of the explicit solution of the density matrix are required for predictions, in the form of $\partial_T |v|$ and  $\partial_T \phi$ analytically known only in weakly coupled environments.

In this work, our goal is to facilitate theory predictions and to elevate the utility of the SLD to a larger class of systems. We develop a \emph{novel method to construct the SLD directly from the master equation}, expressed in terms of accessible expectation values, without requiring the explicit form of the density matrix \(\hat\rho(\theta)\) itself. Our approach leverages the structure of the open-system generator, whether time-local or non-Markovian, and manages to systematically propagate the SLD alongside the dynamical evolution. This formulation makes it possible to \emph{identify optimal observables} at arbitrary times, in particular in the \emph{transient regime}, where the system has not reached a steady state and the full density matrix is often not analytically tractable.

By disentangling the determination of the SLD from the solution of \(\hat\rho(t)\), our method enables metrological optimization in a broad class of experimentally relevant situations, including \emph{driven systems}, \emph{non-equilibrium dynamics}, and \emph{strongly coupled probes}. As such, it provides a valuable tool in the theoretical arsenal of quantum metrology.

This paper is structured as follows. In \cref{sec:form} we introduce our formalism to determine the SLD from the master equation. \Cref{sec:eqqbm} applies the formalism to thermometry in the equilibrium steady state of quantum Brownian motion with a harmonic potential, before constructing in \cref{sec:noneqgauss} the SLD for both temperature and relaxation rate in a non-equilibrium scenario based on a squeezed Gaussian state. \Cref{sec:sum} briefly summarizes the study.

\section{Formalism}
\label{sec:form}
In this section, we present a method for the explicit construction of the symmetric logarithmic derivative (SLD) $\hat \Lambda_\theta$, corresponding to a given environmental parameter $\theta$, expressed in terms of accessible expectation values.

{\subsection{General Construction}}
\label{sec:gencond}
We begin by considering a general master equation governing the evolution of the reduced density matrix $\rho(t,\theta)$ of an open quantum system,
\begin{align}
    \partial_t \hat \rho(t,\theta) = \hat{\mathcal{L}}[\theta] \hat \rho(t,\theta), \label{eq:genmaster}
\end{align}
where $\hat{\mathcal{L}}$ denotes a linear superoperator, which may depend explicitly on the environmental parameter $\theta$. Our formulation does not require the dynamics to be Markovian or weakly coupled, only that the master equation is linear in $\rho$.

The central idea of our approach is to exploit Schwarz's theorem on the symmetry of mixed partial derivatives for twice differentiable functions. This allows us to relate the temporal derivative in the master equation~\eqref{eq:genmaster} to the parameter derivative appearing in the definition of the SLD via
\begin{align}
    \partial_\theta \left( \partial_t \hat \rho(t,\theta) \right) = \partial_t \left( \partial_\theta \hat\rho(t,\theta) \right). \label{eq:identify}
\end{align}
This identity serves as the bridge between the dynamics and the sensitivity operator.

For Markovian evolution, it is known that differentiability of the initial state guarantees differentiability of the evolved state~\cite{mora_regularity_2013}. In general, twice differentiability requires additional smoothness conditions on the generator $\mathcal{L}$, which must be verified on a case-by-case basis. For instance, an explicit analytic treatment of the Caldeira-Leggett master equation~\cite{homa_positivity_2019} confirms that twice differentiability holds for quantum Brownian motion of a point particle in an Ohmic environment \cite{breuer2002theory}.

To make the construction of the SLD concrete, let us expand the operator $\hat \Lambda$ into a basis of Hermitian operators $\hat A_i$
\begin{align}
    \hat \Lambda_\theta = \sum_i c^{(i)}_\theta(t) \hat A_i, \label{eq:ansatzsld}
\end{align}
In practice, we will not have full access to the complete set of basis operators but are limited to the experimentally realizable subset (see discussion in \cref{sec:solv}). That is, the goal is to construct the operator of optimal sensitivity within the set of available operators $A_i$.

While operators in the Schr\"odinger picture are time independent, the SLD inherits a time dependence from its definition \cref{eq:SLDdef}. The change of the reduced density matrix $\hat \rho(t,\theta)$ in time implies that the observable of optimal sensitivity must also change in time. This
time dependence rests solely in the expansion coefficients $c_\theta^{(i)}(t)$ introduced in \cref{eq:ansatzsld}.

In order to convert operator-valued equalities into scalar equations to determine the values of the $c_\theta^{(i)}$, we multiply both sides of Eq.~\eqref{eq:identify} by a test operator $\hat A_j$. Taking the trace leads to
\begin{align}
    \mathrm{Tr}\left[ \partial_\theta \partial_t \hat\rho(t,\theta) \, \hat A_j \right] = \mathrm{Tr}\left[ \partial_t \partial_\theta \hat\rho(t,\theta) \, \hat A_j \right]. \label{eq:convexpval}
\end{align}
Upon substituting the SLD expansion~\eqref{eq:ansatzsld} and computing the partial derivatives, the resulting expression decomposes into three terms (for a detailed derivation see \cref{sec:explicitsld}). One term is independent of the coefficients $c^{(i)}_\theta$ and linear in $\hat A_j$. It will be denoted $D_j$. The second term depends linearly on $c^{(i)}_\theta$ and involves operator bilinears, denoted $M_{ji}$. The third term involves another bilinear $G_{ji}$ and contains the temporal derivatives of the expansion coefficients $\dot c^{(i)}_\theta$. I.e. we are facing a system of linear ordinary differential equations for the $c^{(i)}_\theta$'s, which reads
\begin{align}
    \sum_i \Big\{ G_{ji}(t) \dot c^{(i)}_\theta(t) + M_{ji}(t) \, c^{(i)}_\theta(t) \Big\} = D_j(t)\label{eq:linsysc}.
\end{align}
As we will show in more detail in the following subsection, the system matrix $M$ encodes the full information of the coupling between probe and bath, and thus how the probe records environmental information. Since the system matrix is expressed solely in terms of expectation values of the operators $\langle A_i\rangle$ we will have traded the need to know the explicit form of the density matrix for a knowledge of (accessible) expectation values. On the other hand, the mixing matrix $G$ only makes reference to the expectation values of basis operators and does not involve the generators of the master equation.

Let us first consider \cref{eq:linsysc} in thermal equilibrium, characterized by $\partial_t \hat \rho=0$. Since here the density matrix and thus expectation values of operators become stationary, the SLD too ceases to evolve and the above equation simplifies to 
\begin{align}
    \sum_i M_{ji} \, c^{(i)}_\theta = D_j \label{eq:linsysc2}.
\end{align}
I.e. the time independent SLD of a fully thermalized probe can be constructed from the solution of a set of linear equations, furnished by the system matrix $M_{ij}$, which itself depends on the stationary expectation values $\langle A_i\rangle$  of the accessible observables.

The situation differs out of equilibrium. Here we must in general solve the set of ODEs in \cref{eq:linsysc}. It is a reminder that the form of the SLD not only depends on the current state of the system, but on the shared history of the probe and the surrounding environment. The memory of this history is naturally lost as thermal equilibrium is achieved, where the SLD reduces to the solution of the stationary limit \cref{eq:linsysc2}, a time local quantity.

We will consider the simplest scenario, where the probe is initially fully uncorrelated with the environment. In that state, the probe cannot contain any knowledge of the environment and the QFI is zero. Thus we will set vanishing initial conditions for the expansion coefficients $c^{(i)}_\theta=0$.

Interestingly it turns out (see the following \cref{sec:explicitsld}) that the matrix $G_{ij}$ appearing in the defining \cref{eq:linsysc} is nothing but the covariance matrix of the operator basis $\hat A_i$. As long as strictly independent $\hat A_i$ are chosen $G_{ji}$ will be non-degenerate and thus invertible, so that the SLD can be normalized with the intrinsic (co)-variance of the chosen basis set. Knowledge of the matrix $G$ also allows for a direct computation of the QFI via ${\cal F}_\theta(t) = \sum_{ij} c^{(i)}_\theta(t) G_{ij}(t) c^{(j)}_\theta(t)$.\\

{\subsection{Explicit thermometry SLD from the Lindblad master equation}}
\label{sec:explicitsld}
{The previous section described the general explicit construction of an SLD, which only relied on the linearity of the master equation and the choice of operators in \cref{eq:ansatzsld}. Here we derive the explicit form of the system matrix $M_{ij}$, the matrix $G_{ij}$ and the RHS vector $D_j$ for the optimal observable for measurements of an environment variable $\theta$ in the context of a Markovian Lindblad equation given as
\begin{equation}
    \partial_t\hat{\rho}=-i\comm{\hat{H}}{\hat{\rho}}+\sum_k\gamma_k\qty(\hat{L}_k\hat{\rho}\hat{L}_k^\dagger-\frac{1}{2}\acomm{\hat{L}_k^\dagger\hat{L}_k}{\hat{\rho}})\label{eq:Lindblad}
\end{equation}
Without loss of generality we assume that the relaxation rate parameters $\gamma_k$ are independent of $\theta$.}

Our strategy is to connect \cref{eq:Lindblad} and the defining equation for the thermometry SLD in \cref{eq:SLDdef}. To this end let us apply the time derivative to the latter, which yields 
\begin{widetext}
\begin{equation}
    \partial_t\partial_{\theta}\hat{\rho}=\frac{1}{2}\acomm{\hat{\Lambda}}{\partial_t\hat{\rho}}-\partial_t\hat{\rho}\expval{\Lambda}-\hat{\rho}\Tr[\partial_t\hat{\rho}\hat{\Lambda}] + \frac{1}{2}\acomm{\partial_t{ \hat{\Lambda}}}{\hat{\rho}} -\hat{\rho}\expval{\partial_t \hat{\Lambda}} \label{eq:dtLambda}.
\end{equation}
\end{widetext}
As next step, we must take the temperature derivative of the master equation \cref{eq:Lindblad} which leads to the following expression
\begin{widetext}
\begin{align}
    \partial_{\theta}\partial_t\hat{\rho}&=-i\comm{\partial_{\theta}\hat{H}}{\hat{\rho}}-i\comm{\hat{H}}{\partial_{\theta}\hat{\rho}}+\sum_k\gamma_k\qty(\partial_{\theta}\hat{L}_k\hat{\rho}\hat{L}_k^\dagger-\frac{1}{2}\acomm{\hat{L}_k^\dagger\partial_{\theta}\hat{L}_k}{\hat{\rho}})\nonumber\\
    &+\sum_k\gamma_k\qty(\hat{L}_k\hat{\rho}\partial_{\theta}\hat{L}_k^\dagger-\frac{1}{2}\acomm{\partial_{\theta}\hat{L}_k^\dagger\hat{L}_k}{\hat{\rho}})+\sum_k\gamma_k\qty(\hat{L}_k\partial_{\theta}\hat{\rho}\hat{L}_k^\dagger-\frac{1}{2}\acomm{\hat{L}_k^\dagger\hat{L}_k}{\partial_{\theta}\hat{\rho}})\label{eq:dTLindblad}.
\end{align}
\end{widetext}
After replacing the remaining time derivatives in \cref{eq:dtLambda} via the master equation \cref{eq:Lindblad} we can deploy Schwarz' theorem and equate \cref{eq:dTLindblad} and \cref{eq:dtLambda}. Rearranging terms that depend explicitly on $\Lambda$ on the left and those terms that do not on the right we obtain
\begin{widetext}
\begin{align}
    \nonumber & \frac{1}{2}\acomm{\partial_t{ \hat{\Lambda}}}{\hat{\rho}} -\hat{\rho}\expval{\partial_t \hat{\Lambda}} + \frac{i}{2}\acomm{\hat{\rho}}{\comm{\hat{\Lambda}}{\hat{H}}}-i\hat{\rho}\expval{\comm{\hat{\Lambda}}{\hat{H}}}+\hat{\rho}\sum_k\gamma_k\qty(\expval{\hat{L}_k^\dagger\hat{\Lambda}\hat{L}_k}-\frac{1}{2}\expval{\acomm{\hat{\Lambda}}{\hat{L}_k^\dagger\hat{L}_k}})\\
    \nonumber &+\frac{1}{2}\sum_k\gamma_k\qty(\hat{L}_k\hat{\Lambda}\hat{\rho}\hat{L}_k^\dagger-\frac{1}{2}\hat{L}_k^\dagger\hat{L}_k\hat{\Lambda}\hat{\rho}+\text{h.c.})-\frac{1}{2}\sum_k\gamma_k\qty(\hat{\Lambda}\hat{L}_k\hat{\rho}\hat{L}_k^\dagger-\frac{1}{2}\hat{\Lambda}\hat{L}_k^\dagger\hat{L}_k\hat{\rho}+\text{h.c.})\\
    &=i\comm{\partial_{\theta}\hat{H}}{\hat{\rho}}-\sum_k\gamma_k\qty(\partial_{\theta}\hat{L}_k\hat{\rho}\hat{L}_k^\dagger-\frac{1}{2}\acomm{\hat{L}_k^\dagger\partial_{\theta}\hat{L}_k}{\hat{\rho}}+\text{h.c.}).
\end{align}
\end{widetext}
Let us now introduce a sufficiently expressive set of operators that is closed under the generators of the master equation, as discussed in the preceding subsection. Then with $n$ Hermitian operators $\hat{A}_i$, we express $\hat{\Lambda}=\sum_ic_i\hat{A}_i$ with $c_i$ c-numbers so that
\begin{widetext}
\begin{align}
   \nonumber  & \sum_i\dot c_i\left[ \frac{1}{2}\acomm{\hat A_i}{\hat{\rho}} -\hat{\rho}\expval{A_i} \right]  \\
    \nonumber &+
   \sum_ic_i\left[\frac{i}{2}\acomm{\hat{\rho}}{\comm{\hat{A}_i}{\hat{H}}}-i\hat{\rho}\expval{\comm{\hat{A}_i}{\hat{H}}}+\hat{\rho}\sum_k\gamma_k\qty(\expval{\hat{L}_k^\dagger\hat{A}_i\hat{L}_k}-\frac{1}{2}\expval{\acomm{\hat{A}_i}{\hat{L}_k^\dagger\hat{L}_k}})\right.\\
   \nonumber  &\left.+\frac{1}{2}\sum_k\gamma_k\qty(\hat{L}_k\hat{A}_i\hat{\rho}\hat{L}_k^\dagger-\frac{1}{2}\hat{L}_k^\dagger\hat{L}_k\hat{A}_i\hat{\rho}+\text{h.c.})-\frac{1}{2}\sum_k\gamma_k\qty(\hat{A}_i\hat{L}_k\hat{\rho}\hat{L}_k^\dagger-\frac{1}{2}\hat{A}_i\hat{L}_k^\dagger\hat{L}_k\hat{\rho}+\text{h.c.})\right]\\
    &=i\comm{\partial_{\theta}\hat{H}}{\hat{\rho}}-\sum_k\gamma_k\qty(\partial_{\theta}\hat{L}_k\hat{\rho}\hat{L}_k^\dagger-\frac{1}{2}\acomm{\hat{L}_k^\dagger\partial_{\theta}\hat{L}_k}{\hat{\rho}}+\text{h.c.})
\end{align}
\end{widetext}
The above equation is still an operator equation. To efficiently extract the values of the $c_i$'s we thus multiply both LHS and RHS with the same test operators denoted by $\hat A_j$ and take the trace. This allows us to convert expressions containing the density matrix into expectation values. The c-number equation then reads
\begin{widetext}
\begin{align}
   \nonumber &\sum_i\dot c_i\left[ \frac{1}{2}\expval{ \acomm{\hat A_i}{\hat A_j}} -\expval{\hat A_i} \expval{\hat A_j} \right] \\
    \nonumber &+ \sum_ic_i\left[\frac{i}{2}\expval{\acomm{\comm{\hat{A}_i}{\hat{H}}}{\hat{A}_j}}-i\expval{A_j}\expval{\comm{\hat{A}_i}{\hat{H}}}+\expval{A_j}\sum_k\gamma_k\qty(\expval{\hat{L}_k^\dagger\hat{A}_i\hat{L}_k}-\frac{1}{2}\expval{\acomm{\hat{A}_i}{\hat{L}_k^\dagger\hat{L}_k}})\right.\\
   \nonumber &\left.+\frac{1}{2}\sum_k\gamma_k\qty(\expval{\hat{L}_k^\dagger\hat{A}_j\comm{\hat{L}_k}{\hat{A}_i}}-\frac{1}{2}\expval{\hat{A}_j\comm{\hat{L}_k^\dagger\hat{L}_k}{\hat{A}_i}}+\text{c.c.})\right]\\
    &=i\expval{\comm{\hat{A}_j}{\partial_{\theta}\hat{H}}}-\sum_k\gamma_k\qty(\expval{\hat{L}_k^\dagger\hat{A}_j\partial_{\theta}\hat{L}_k}-\frac{1}{2}\expval{\acomm{\hat{A}_j}{\hat{L}_k^\dagger\partial_{\theta}\hat{L}_k}}+\text{c.c.})
\end{align}
\end{widetext}
Extracting the dependence on the expansion coefficients $c_i$ up front shows that we are dealing with a system of linear ordinary differential equations $\sum_j \Big\{  G_{ij}\dot c_j + M_{ij}c_j \Big\}=D_i$ where the system matrix takes the explicit form
\begin{widetext}
\begin{align}
    \nonumber M_{ij}&=\frac{i}{2}\expval{\acomm{\hat{A}_j}{\comm{\hat{A}_i}{\hat{H}}}}-i\expval{A_j}\expval{\comm{\hat{A}_i}{\hat{H}}}+\sum_k\gamma_k\qty(\expval{A_j}\expval{\hat{L}_k^\dagger\hat{A}_i\hat{L}_k}-\frac{1}{2}\expval{A_j}\expval{\acomm{\hat{A}_i}{\hat{L}_k^\dagger\hat{L}_k}})\\
    &+\Re{\sum_k\gamma_k\qty(\expval{\hat{L}_k^\dagger\hat{A}_j\comm{\hat{L}_k}{\hat{A}_i}}-\frac{1}{2}\expval{\hat{A}_j\comm{\hat{L}_k^\dagger\hat{L}_k}{\hat{A}_i}})}\label{eq:MijLindblad},
\end{align}
\end{widetext}
the mixing matrix of the derivatives reads
\begin{align}
    \nonumber G_{ij}&=\frac{1}{2}\expval{ \acomm{\hat A_i}{\hat A_j}} -\expval{ A_i} \expval{ A_j}\label{eq:GijLindblad}.
\end{align}
Note that $G_{ij}$ does not make any reference to $\theta$ derivatives of the generators of the master equation. In fact $G_{ij}$ is nothing but the covariance matrix of the quantum operators $\hat A_i$ underlying our expansion of the SLD in \cref{eq:ansatzsld}.

The $\Lambda$-independent RHS becomes
\begin{align}
    \nonumber D_j=&i\expval{\comm{\hat{A}_j}{\partial_{\theta}\hat{H}}} -2{\rm Re}\left\{\sum_k\gamma_k\left(\expval{\hat{L}_k^\dagger\hat{A}_j\partial_{\theta}\hat{L}_k} \right.\right.\\
    &\left.\left.-\frac{1}{2}\expval{\acomm{\hat{A}_j}{\hat{L}_k^\dagger\partial_{\theta}\hat{L}_k}}\right)\right\}.
\end{align}
This concludes the construction of the left- and right-hand side of the linear system of ordinary differential equations whose solution furnishes the expansion coefficients $c_i$ of the SLD.\\

{\subsection{Conditions for Solvability}}
\label{sec:solv}

The defining equation \cref{eq:linsysc} constitutes a linear ordinary differential equation. The associated initial value problem is well-posed under relatively weak conditions. By choosing strictly independent $\hat A_i$, in order for $G_{ji}$ to be non-degenerate, we only need to require continuity in $G^{-1}M$ and $G^{-1}D$ and thus continuity in $M(t)$ and $D(t)$. 

In order to guarantee the convergence to the correct stationary solution however, which is obtained from the solution of the linear system of equations \cref{eq:linsysc2}, we find that stronger conditions apply. In the following we will focus on the existence of this solution of the stationary limit.

For the linear system \cref{eq:linsysc2} to have a unique solution, the system matrix $M$ must be non-degenerate. Since the SLD is an observable and thus a Hermitian operator acting on vectors in the subsystem Hilbert space, it is guaranteed that if a complete basis of operators $\{\hat A_i\}$ is available, the SLD can be expressed in it.

In practice however only a subset ${\cal A} = {\rm span}_N\{\hat A_i\}$ of $N$ observables is accessible and we must ascertain whether the matrix $M$ evaluated on ${\cal A}$ is invertible. There are two different scenarios to consider. If we use too narrow a set of observables, then the system of equations may not have a solution at all. On the other hand if we add observables that are not independent of those already included there may be an infinite number of solutions reflecting the ambiguity in expressing the SLD.

While we are unable to give general criteria for solvability, we report here on relevant heuristic observations. 

Let us consider the rows of the matrix $M$ as maps from the space of operators ${\cal M}: {\cal A} \to \mathbb{R}^N $ into Euclidean space, i.e. $M_{ij}=M_j(\hat O)$. Then our goal is to ascertain that there do not exist any operator $\hat O = \sum_i c_i \hat A_i$ that annihilates all $M_j$. This setup tells us that the commutator and anticommutator structures appearing e.g. in the concrete example of \cref{eq:MijLindblad} are key to understanding whether ${\cal M}$ has a trivial kernel. 

When we wish to address the absence of a solution, we may generate relevant operators to include in ${\cal A}$ by evaluating the commutators and anticommutators appearing in the definition of $M$. A concrete example is discussed in \cref{sec:eqqbm} where the inclusion of the operator ${\hat x,\hat p}$ is key to obtain a well-defined SLD. If we instead started by excluding the anticommutator and only consider $\hat x^2$ and $\hat p^2$, already the commutator with the Hamiltonian $[\hat H, \hat A_j]$ in \cref{sec:eqqbm} will generate expressions involving  $\{\hat x,\hat p\}$. Conversely when $\{\hat x,\hat p\}$ is included in the set of operators the (anti-)commutator terms do not generate additional operators, i.e. the set of operators remains closed under the structures arising within $M$.

This argument extends to cases with operators in subspaces of the full Hilbert space. Take as example a probe system consisting of two qubits. If we start with observables which only make reference to one of the qubits, (anti-)commutator terms with the system Hamiltonian will in general generate expressions that include operators from the second qubit.

The case of an overspecified set of operators is easier to handle, as one can systematically remove offending observables. A relevant example is the inclusion of spin-dependent observables when the underlying master equation does not distinguish different spin states. Using instead a single spin-averaged observable will remove the ambiguity.

Due to the generality of our construction strategy for the SLD, it applies to a wide range of master equations, leading to different (anti-)commutator structures for the system matrix $M$. In the absence of a general criterion for solvability, an inspection of the concrete structure of $M$ on a case-by-case basis therefore remains necessary.\\

\section{Equilibrium probe thermometry in Quantum Brownian Motion}
\label{sec:eqqbm}

In this section we verify our formalism by constructing the optimally sensitive observable for temperature measurements in the steady state of quantum Ornstein-Uhlenbeck motion \cite{uhlenbeck1930theory,del2017stochastic}, i.e. quantum Brownian motion in the presence of a harmonic potential \cite{breuer2002theory}. This motion describes the dynamics of a confined and heavy point particle impurity in an Ohmic environment and leads to a stationary state at late time with time-independent expectation values. We use the Caldeira-Leggett (CL) master equation \cite{caldeira_path_1983,caldeira_influence_1985} in its original, non-Lindblad form to follow the evolution of the underlying reduced density matrix of the impurity. The key result is that our approach reproduces the known solution for the SLD, obtained from the explicit solution of the master equation in the late-time limit.

Making use solely of the master equation, let us inspect it in more detail. Denoting the single-particle Hamiltonian by $\hat H=\frac{\hat p^2}{2m}+\frac{1}{2}m\omega^2\hat x^2 = b\hat p^2 +c \hat x^2$, we have
\begin{align}
&\partial_t \hat\rho(t,T) =\label{eq:CLmaster}\\&-i [\hat H, \hat\rho] - i\gamma [\hat x,\{\hat p,\hat\rho\}]-2\gamma m T[ \hat x,[\hat x,\hat \rho]],\nonumber 
\end{align}
where $[\cdot,\cdot]$ and $\{\cdot,\cdot\}$ denote the commutator and anticommutator respectively. The subsystem parameters are the mass $m$ and oscillation frequency $\omega$, while the environment is specified by its temperature $T$ and the coupling to the probe in the form of the relaxation rate $\gamma$. 

In the absence of inner structure the two Hermitian operators that characterize all motion of the probe are the position $\hat x$ and momentum $\hat p$ operators. We thus consider powers of these two operators as basis for the operator expansion of \cref{eq:ansatzsld}. Since one- and two-point functions are the most easily accessible quantities in experiment, our truncated basis reads explicitly
\begin{align}
\hat A_j \in \{ \hat x, \hat p, \hat x^2, \hat p^2, \{\hat x,\hat p\} \}, \label{eq:opbasis}
\end{align} 
where the last term denotes the anticommutator of position and momentum.

The key relation \cref{eq:identify} requires us to take the derivative w.r.t. the environment parameter of interest, here $T$, of the master equation. Whenever $\partial_T$ is applied to the density matrix, $\partial_T\hat\rho$ will be replaced by the definition of the SLD \cref{eq:SLDdef}. Temperature appears explicitly on the r.h.s. of \cref{eq:CLmaster}. In turn after applying $\partial_T$, there remains one single term where no temperature derivative of the density matrix occurs. After converting the operator expressions to expectation values with the help of a test operator $\hat A_j$ via \cref{eq:convexpval} this single term will become $D_j$ in \cref{eq:linsysc2}. The remaining terms depend linearly on $\Lambda$, so that when the SLD too is expanded in an operator basis via \cref{eq:ansatzsld}, they furnish the l.h.s. of \cref{eq:linsysc2}.

The linear system of equations that defines the expansion coefficients $c^{(i)}_T$ of the SLD thus reads
\begin{align}
&M_{ij} =\label{eq:Mij}\\ 
& -\tfrac{1}{2} \left\langle \left\{ \hat{A}_i, \left[ \hat{H}, \hat{A}_j \right] \right\}\right\rangle 
- \tfrac{i}{2} \left\langle \left\{ \hat{A}_i, \left\{ \hat{p} ,\left[ \hat{A}_j, \hat{x} \right] \right\} \right\} \right\rangle \nonumber \\
& - \tfrac{\gamma T}{2 b} \left\langle \left\{ \hat{A}_i, \left[ \hat{x}, \left[ \hat{x}, \hat{A}_j \right] \right] \right\} \right\rangle 
 \nonumber\\&+ \tfrac{i \gamma}{2} \left\langle \left\{ \hat{p}, \left[ \left\{ \hat{A}_i, \hat{A}_j\right\}, \hat{x} \right] \right\}\right\rangle \nonumber\\
& + \tfrac{\gamma T}{2 b} \left\langle \left[ \hat{x}, \left[ \hat x, \left\{ \hat{A}_i, \hat{A}_j \right\}\right] \right] \right\rangle 
- i \left \langle \hat{A}_j \right\rangle \left\langle \left[ \hat{A}_i, \hat{H} \right] \right\rangle \nonumber \\
\nonumber& - i \gamma \left\langle \hat{A}_j \right\rangle \left\langle \left\{ \hat{p}, \left[ \hat{A}_i, \hat{x} \right] \right\} \right\rangle 
\nonumber\\&- \tfrac{\gamma T}{b} \left\langle \hat{A}_j \right\rangle \left\langle \left[ \hat x, \left[ \hat{x}, \hat{A}_i \right] \right] \right\rangle\nonumber
\end{align}
with the constant terms
\begin{align}
    D_j=\frac{\gamma}{b}\left\langle \left[ \hat x,\left[ \hat x, \hat A_j \right] \right]\right\rangle.\label{eq:Dj}
\end{align}

Let us evaluate the entries of the matrix $M$ and the vector $D$ in the late-time steady state, where the system has become Gaussian. For an efficient handling of terms with different powers of the $\hat x$ and $\hat p$ operator we collect them in terms of the Weyl symbols \cite{mccoy_function_1932}
\begin{align}
W(\hat{x}^n\hat{p}^m)=\frac{1}{2^n}\sum_{k=0}^n\binom{n}{k}\hat{x}^{n-k}\hat{p}^m\hat{x}^k.
\end{align}
The expectation value of any Weyl symbol with an odd power of either operator vanishes, such as e.g. 
\begin{align}
 \left \langle W(xp) \right \rangle = \frac{1}{2} \left \langle\left\{ \hat x,\hat p \right\} \right \rangle= 0,\label{eq:Wxp}\\
  \left \langle W(x^3p) \right \rangle = 0, \quad  \left \langle W(xp^3) \right \rangle = 0,\label{eq:Wx3p}
\end{align}
while the only relevant non-vanishing Weyl symbol is
\begin{align}
    \left \langle W(x^2p^2) \right \rangle = \langle x^2\rangle \langle p^2\rangle.
\end{align}
Expectation values of higher powers of the individual operators in the Gaussian state factor in products of lower powers according to Wick's theorem \cite{wick1950evaluation}. The relevant expressions are
\begin{align}
    &\langle x^4\rangle = 3 \langle x^2\rangle^2, \quad \langle \hat p^4\rangle = 3 \langle \hat p^2\rangle^2\\
    &\langle x^2\rangle = \frac{a^2 b}{4c}, \quad  \langle \hat p^2\rangle = \frac{a^2}{4},
\end{align}
where we have introduced the abbreviation $a=\sqrt{4mT}$.

A careful evaluation of all $5\times5$ entries of $M$ yields
\begin{widetext}
\begin{align}
M=&
\left[
\begin{array}{ccccc}
 0 & -2 b \left\langle p^2\right\rangle  & 0 & 0 & 0 \\[10pt]
 2 c \left\langle x^2\right\rangle  & \begin{array}{c} 2 \gamma  \left\langle p^2\right\rangle \\ -\frac{2 \gamma  T}{b} \end{array} & 0 & 0 & 0 \\[4pt]
 0 & 0 & 0 & 2 \gamma  & -8 b \left\langle p^2\right\rangle  \left\langle x^2\right\rangle -2 b \\[4pt]
 0 & 0 & 0 & \begin{array}{c} 8 \gamma  \left\langle p^2\right\rangle ^2\\ -\frac{8 \gamma  T \left\langle p^2\right\rangle }{b}  \end{array}& 8 c \left\langle
   p^2\right\rangle  \left\langle x^2\right\rangle +2 c \\[4pt]
 0 & 0 & \begin{array}{c} 2 b\\+8 c \left\langle x^2\right\rangle ^2 \end{array} & \begin{array}{c} -8 b \left\langle p^2\right\rangle ^2\\ -2 c  \end{array} & \begin{array}{c} -\frac{8 \gamma  T \left\langle
   x^2\right\rangle }{b}-2 \gamma\\ +8 \gamma  \left\langle p^2\right\rangle  \left\langle x^2\right\rangle \end{array}  \\
\end{array}
\right]\\
\nonumber=& \begin{bmatrix}
0 & -\frac{b a^2}{2} & 0 & 0 & 0 \\
\frac{a^2 b}{2} & -\frac{\gamma a^2}{2} & 0 & 0 & 0 \\
0 & 0 & 0 & 2 \gamma & -\dfrac{a^4 b^2}{2 c} - 2 b \\
0 & 0 & 0 & -\dfrac{\gamma a^4}{2} & \frac{a^4 b}{2} + 2 c \\
0 & 0 & \frac{a^4 b^2}{2 c} + 2 b & -\frac{a^4 b}{2} - 2 c & -\frac{\gamma a^4 b}{2 c} - 2 \gamma
\end{bmatrix},
\end{align}
\end{widetext}
where after the second equality, the expectation values have been replaced by their explicit values in the late-time limit, in terms of the model parameters. The constant term features a single non-vanishing entry and reads
\begin{align}
D_T = \left[ 0,0,0, - 2\frac{\gamma}{b},0\right].
\end{align}
The entries are ordered according to the ordering in \cref{eq:opbasis}, i.e. the entry $M_{00} = M_{xx}$ and the entry $M_{44}= M_{ \{x,p\} \{x,p\} }$.

Interesting structure emerges in the system of linear equations. Since the Caldeira-Leggett equation is not symmetric in position and momentum the matrix $M$ is not necessarily symmetric. In addition we find that the terms associated with a single power of the operators (upper-left $2\times2$ entries) factorize from the contributions related to products of operators (lower-right $3\times 3$ entries). Due to the fact that the constant terms vanish except for the entry $D_{p^2}$, we obtain as solution
\begin{align}
    c^{(x)}_T &= 0, \label{eq:SLDTequilcoeff}\\
    \nonumber c^{(p)}_T &= 0,\\
    \nonumber c^{(x^2)}_T &= \frac{4 c \left( a^4 b^2 + 4 (b c + \gamma^2) \right)}{a^8 b^4 - 16 b^2 c^2},\\
    \nonumber c^{(p^2)}_T &= \frac{4}{a^4 b - 4 c},\\
    \nonumber c^{( \{x,p\})}_T &= \frac{16 c \gamma}{a^8 b^3 - 16 b c^2}.
\end{align}
An explicit implementation of the system matrix and of the expansion coefficients obtained from it can be found in the accompanying \texttt{Mathematica} scripts, available open-access at the Zenodo repository \cite{rothkopfZ:2025}.

Let us compare the expressions obtained here with the known non-vanishing solutions $\bar c^{(x^2)}_T$ and $\bar c^{(p^2)}_T$ obtained in Ref.~\cite{mehboudi_using_2019} from the explicit Gaussian density matrix in the late time limit
\begin{align}
    \bar c^{(x^2)}_T &= \frac{4 c \left( a^4 b^2 + 4 b c  \right)}{a^8 b^4 - 16 b^2 c^2},\\
    \bar c^{(p^2)}_T &= \frac{4}{a^4 b - 4 c}.
\end{align}
We find that in the Markovian approximation underlying the CL equation $\gamma \ll 2\pi T$ we can estimate $a^4b^2 = 4 T^2 \gg  4\gamma^2 $ and thus our solution agrees with the known result.

Our approach however reveals an important detail. For the SLD to be well defined, i.e. to obtain a consistent solution of the linear system of equations the operator basis must include $\{\hat x,\hat p\}$, even though the expectation value of that operator vanishes in the late time limit. The presence of $\gamma$ in our solution serves as a reminder that the system must be able to approach the late-time steady state, which is only possible with a finite value of $\gamma$, while the actual value of $\gamma$ does not play a role for the optimal observable itself
\begin{align}
    \hat \Lambda_T =  c^{(x^2)}_T \hat x^2 + c^{(p^2)}_T \hat p^2.
\end{align}

This result is encouraging, as we managed to construct an explicit expression for the SLD, based only on our knowledge of the master equation and which reproduces the known equilibrium result.\\

\section{Non-equilibrium metrology for a squeezed gaussian state}
\label{sec:noneqgauss}

In this section we apply our approach to non-equilibrium metrology, i.e. we consider a probe that is not yet in equilibrium with its environment. 

Our goal is to derive not only the optimally sensitive observable for the determination of the temperature of the environment, but also for the relaxation rate $\gamma$. The latter is not directly accessible in the late-time steady state, where most information about the initial conditions, and thus the approach to equilibrium has been lost. {Renormalization of the probe Hamiltonian, manifest in an in-medium frequency shift in the potential, does encode a remnant of the bath interactions that survives into the late-time steady state but is not equivalent to the relaxation rate. 

In an Ohmic bath with spectral density $J(\omega)$ the relaxation rate is defined as $\gamma=\lim_{\omega\to0} J(\omega)/m\omega$ while $\Delta \omega \propto \int d\omega J(\omega)/m\omega$.} As a concrete realization, we consider a squeezed Gaussian state in which $\langle \hat x\rangle=0$, $\langle \hat p \rangle =0 $ but $\langle \{ \hat x, \hat p\}\rangle\neq 0$. {We note that while an explicit solution of the CL master equation is possible using the Feynman-Vernon influence functional, our explicit construction of the SLD in the transient regime in  the following section does not rely on that solution.}

\subsection{Non-equilibrium Thermometry}

For the non-equilibrium SLD for thermometry we must solve the defining differential equation \cref{eq:linsysc}. In contrast to the equilibrium setting, we now need access to the covariance matrix $G_{ij}=\tfrac12\langle\{\hat A_i,\hat A_j\}\rangle
-\langle\hat A_i\rangle\langle\hat A_j\rangle$ of our operator basis of choice $\hat A=(\hat x,\hat p,\hat x^2,\hat p^2,\{\hat x,\hat p\})$. 

We must keep in mind that for a squeezed Gaussian state, mixed expectation values do not necessarily vanish but they factorize according to Wick's theorem \cite{wick1950evaluation}, such as 
\begin{align}
    &\langle W(x^3 p) \rangle = \frac{3}{2} \langle x^2 \rangle \langle \left\{ \hat x,\hat p\right\} \rangle,\label{eq:Wx3pSqueezed}\\
    & \left \langle W(x^2p^2) \right \rangle = \langle x^2\rangle \langle p^2\rangle + \frac{1}{2} \left \langle \left\{ \hat x,\hat p\right\} \right \rangle^2.
\end{align}

Let us first construct the general covariance matrix $G$ using expectation values and cumulants 
\begin{align}
&\sigma_x=\langle\hat x^2\rangle-\langle\hat x\rangle^2\\
&\sigma_p=\langle\hat p^2\rangle-\langle\hat p\rangle^2\\
&\sigma_{xp}=\tfrac{1}{2}\langle\{\hat x,\hat p\}\rangle
  -\langle\hat x\rangle\langle\hat p\rangle,
\end{align}
which takes on the following form
\begin{widetext}
\begin{align}
G=
\begin{bmatrix}
\sigma_x & \sigma_{xp} & 2\langle\hat x\rangle\sigma_x
  & 2\langle\hat p\rangle\sigma_{xp}
  & 2\left(\langle\hat x\rangle\sigma_{xp}+\langle\hat p\rangle\sigma_x\right)\\[1.5mm]
\sigma_{xp} & \sigma_p &  2\langle\hat x\rangle\sigma_{xp}
  & 2\langle\hat p\rangle\sigma_p
  & 2\left(\langle\hat p\rangle\sigma_{xp}+\langle\hat x\rangle\sigma_p\right)\\[1.5mm]
 2\langle\hat x\rangle\sigma_x &  2\langle\hat x\rangle\sigma_{xp}
  & 2\sigma_x^2+4\langle\hat x\rangle^2\sigma_x
  &  4\langle\hat x\rangle\langle\hat p\rangle\sigma_{xp}
      +2\sigma_{xp}^2-\frac{1}{2}
  & 4\left(\langle\hat x\rangle^2\sigma_{xp}
      +\langle\hat x\rangle\langle\hat p\rangle\sigma_x
      +\sigma_x\sigma_{xp}\right)\\[1.5mm]
 2\langle\hat p\rangle\sigma_{xp} & 2\langle\hat p\rangle\sigma_p
  & 4\langle\hat x\rangle\langle\hat p\rangle\sigma_{xp}
      +2\sigma_{xp}^2-\frac{1}{2}
  & 2\sigma_p^2+4\langle\hat p\rangle^2\sigma_p
  & 4\left(\langle\hat p\rangle^2\sigma_{xp}
      +\langle\hat x\rangle\langle\hat p\rangle\sigma_p
      +\sigma_p\sigma_{xp}\right)\\[1.5mm]
\begin{array}{c} 2(\langle\hat x\rangle\sigma_{xp}\\ +\langle\hat p\rangle\sigma_x)\end{array}
  & \begin{array}{c} 2(\langle\hat p\rangle\sigma_{xp}\\ +\langle\hat x\rangle\sigma_p)\end{array}
  & \begin{array}{c}  4(\langle\hat x\rangle^2\sigma_{xp}
      +\langle\hat x\rangle\langle\hat p\rangle\sigma_x
      \\  +\sigma_x\sigma_{xp})\end{array}
  & \begin{array}{c}  4(\langle\hat p\rangle^2\sigma_{xp}
      +\langle\hat x\rangle\langle\hat p\rangle\sigma_p
      \\+\sigma_p\sigma_{xp})\end{array}
  & \begin{array}{c} 4(\sigma_x\sigma_p+\sigma_{xp}^2\\
  +\langle\hat x\rangle^2\sigma_p+\langle\hat p\rangle^2\sigma_x\\
  +2\langle\hat x\rangle\langle\hat p\rangle\sigma_{xp})+1
  \end{array}
\end{bmatrix},
\label{eq:Gmatrix}
\end{align}
\end{widetext}
For $\langle\hat x\rangle=\langle\hat p\rangle=0$ as chosen in our example here, the linear-quadratic
blocks vanish and $G$ too factorizes into a $2\times2$ and a $3\times3$ block structure. 

Continuing along the lines developed in the previous subsection, we obtain the following explicit expression for the system matrix $M$:

\begin{figure*}[t]
\centering
\includegraphics[scale=0.35]{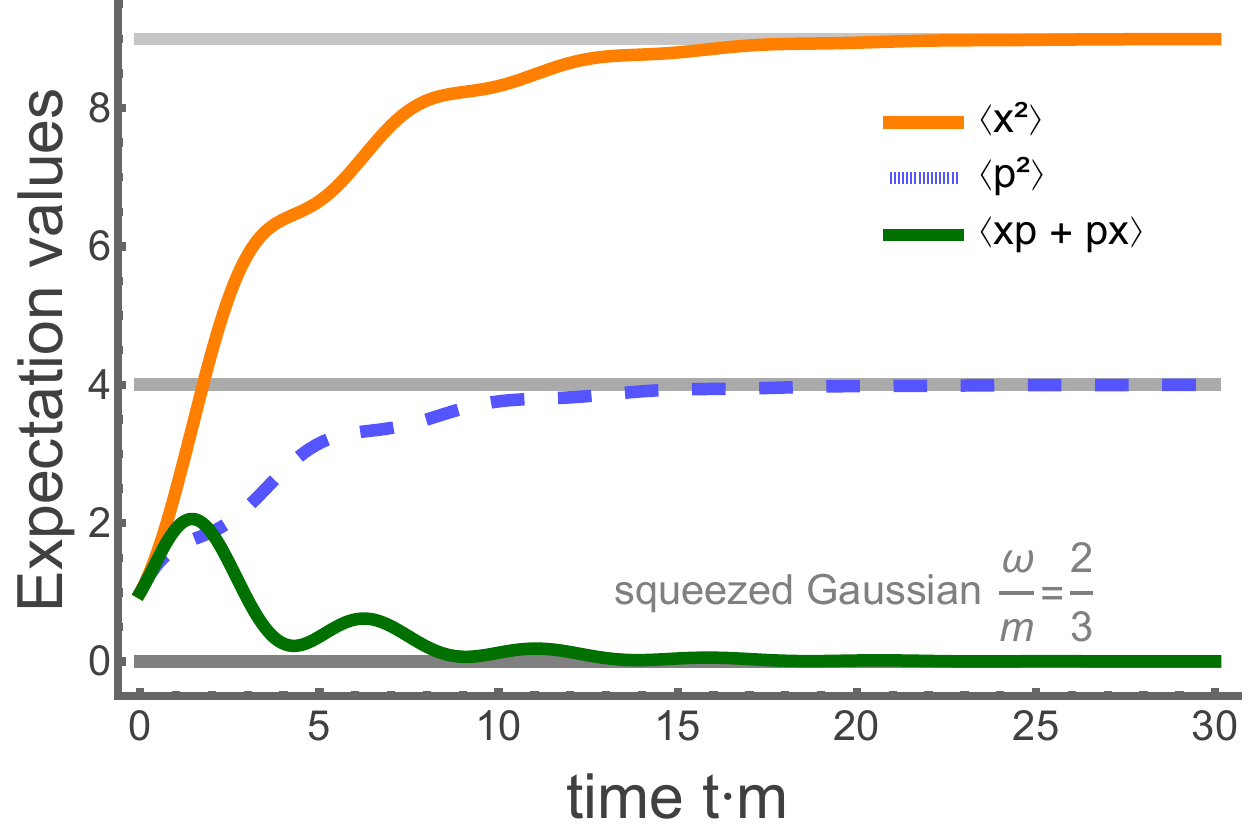}
\includegraphics[scale=0.35]{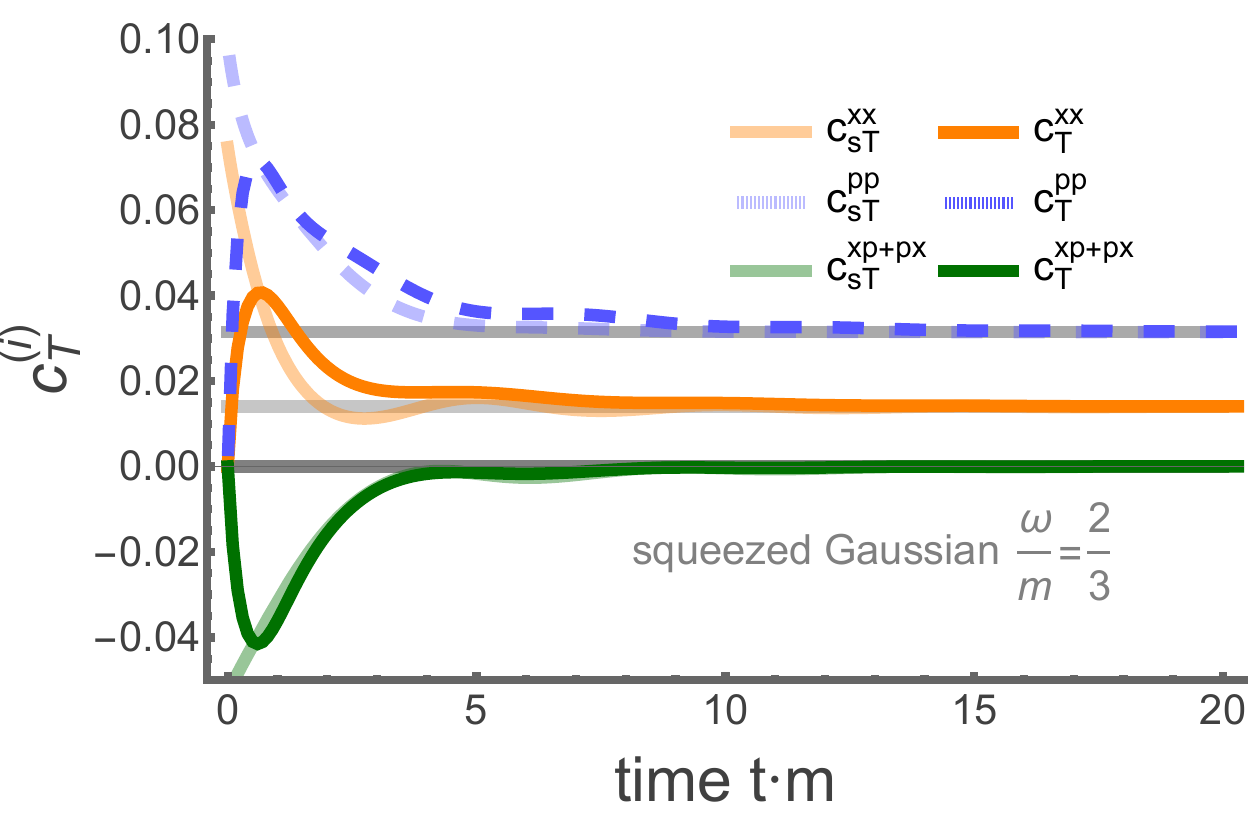}
\includegraphics[scale=0.35]{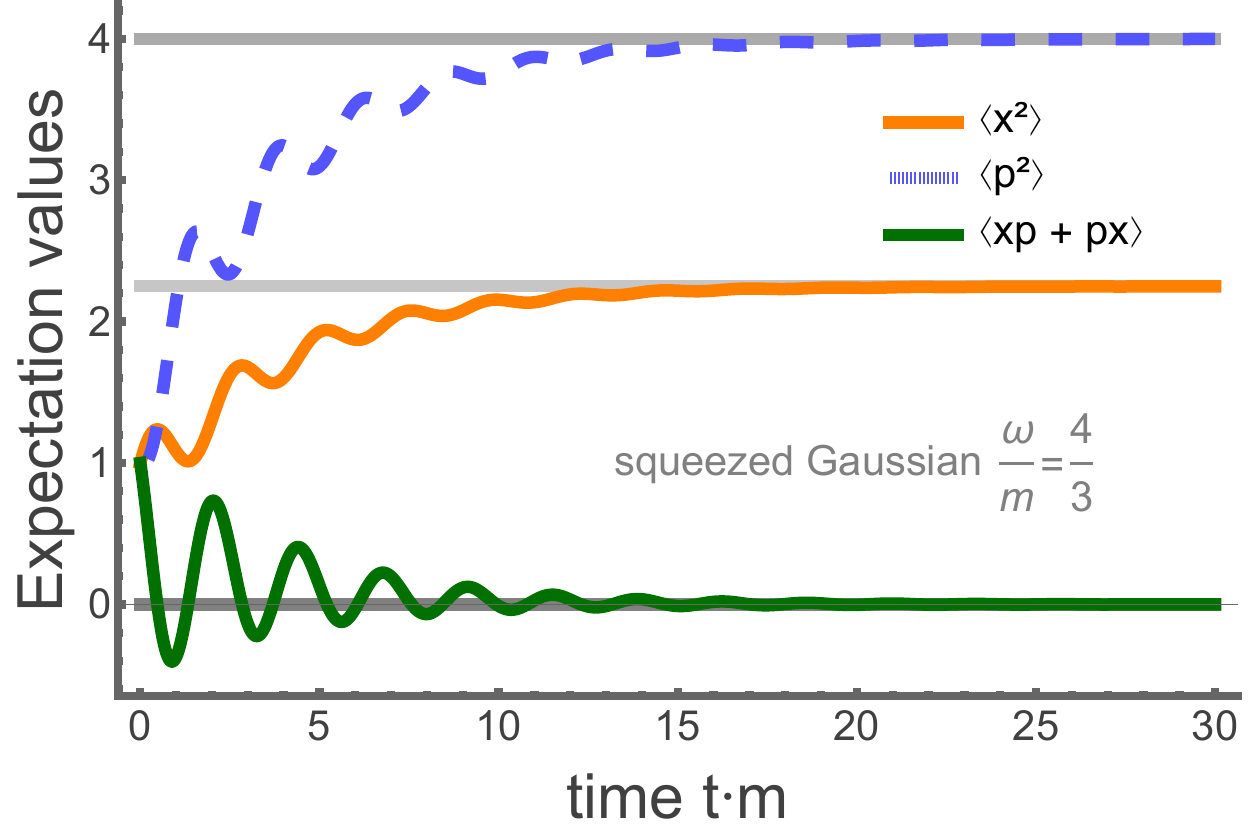}
\includegraphics[scale=0.35]{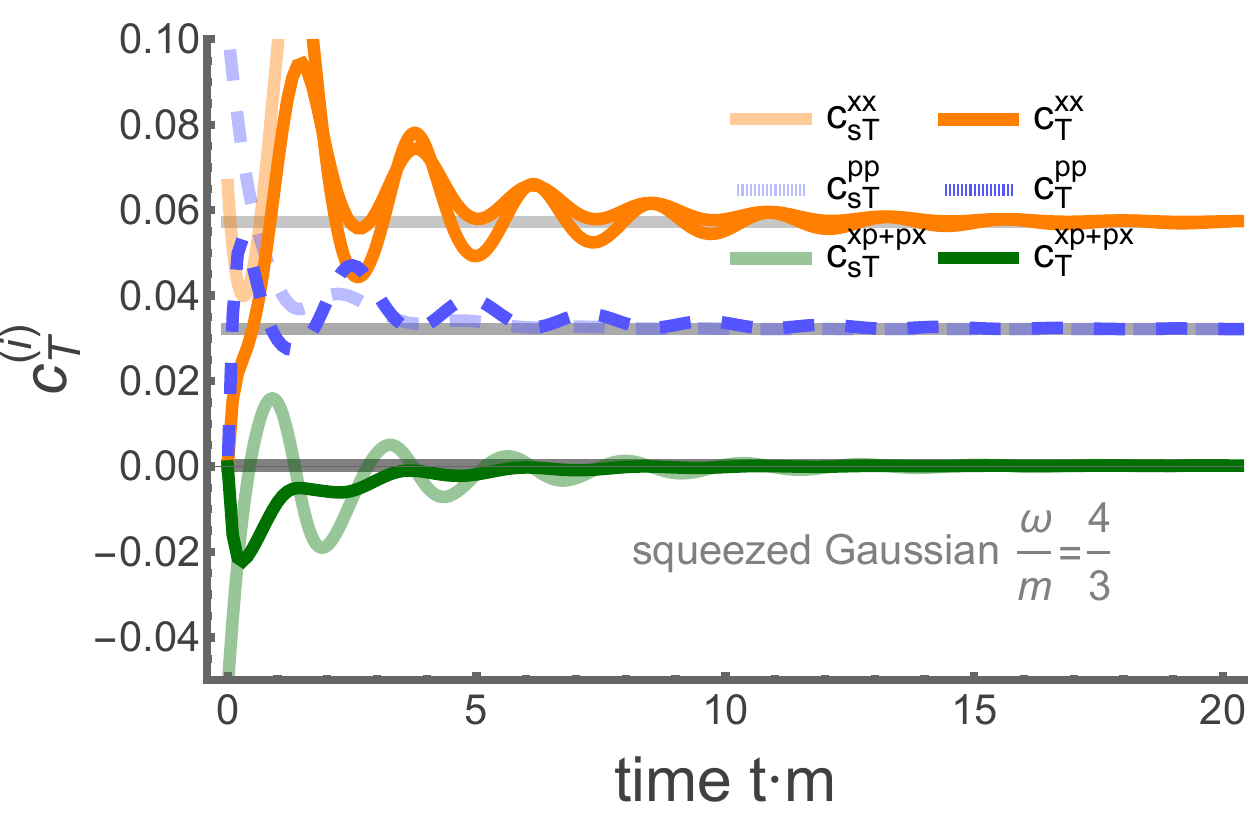}
\caption{Probe particle evolution in a squeezed Gaussian state for (top) $\omega/m=2/3$ and (bottom) $\omega/m=4/3$. Evolution of the expectation values (left) of the relevant observables $\langle x^2 \rangle$ (orange solid), $\langle p^2 \rangle$ (blue dashed), $\langle px + xp \rangle$ (green solid). Their equilibrium values are shown as solid gray lines. (right) Relevant expansion coefficients $c^{(i)}_T$ of the symmetric logarithmic derivative for temperature. Here we compare the coefficients obtained from the solution of the defining differential equation \cref{eq:linsysc} (opaque lines) with the coefficients obtained from evaluating simply the stationary approximation \cref{eq:linsysc2} at the same time step (transparent lines). Note that the true coefficients consistently start from zero, while the static approximation overestimates the sensitivity at early times. The equilibrium solution of the coefficients is also shown as gray solid line.} \label{fig:sqGaussEvol}
\end{figure*}

\begin{widetext}
\begin{equation}
M=\begin{bmatrix}
\begin{array}{c}\scriptstyle- b \langle \{x, p\} \rangle \end{array} &
\begin{array}{c}\scriptstyle-2 b \langle p^2 \rangle \end{array} &
\begin{array}{c}\scriptstyle0\end{array} &
\begin{array}{c}\scriptstyle0\end{array} &
\begin{array}{c}\scriptstyle0\end{array} \\[10pt]

\begin{array}{c}
\scriptstyle 2 c \langle x^2 \rangle \\\scriptstyle
+ \gamma \langle \{x, p\} \rangle
\end{array} &
\begin{array}{c}\scriptstyle
2 \gamma \langle p^2 \rangle - \frac{2 \gamma T}{b} \\\scriptstyle
+ c \langle \{x, p\} \rangle
\end{array} &
\begin{array}{c}\scriptstyle0\end{array} &
\begin{array}{c}\scriptstyle0\end{array} &
\begin{array}{c}\scriptstyle0\end{array} \\[20pt]

\begin{array}{c}\scriptstyle0\end{array} &
\begin{array}{c}\scriptstyle0\end{array} &
\begin{array}{c}\scriptstyle
-4 b \langle x^2 \rangle \langle \{x, p\} \rangle
\end{array} &
\begin{array}{c}\scriptstyle
-4 b \langle p^2 \rangle \langle \{x, p\} \rangle \\\scriptstyle
+ 2 \gamma
\end{array} &
\begin{array}{c}\scriptstyle
-8 b \langle x^2 \rangle \langle p^2 \rangle - 2 b \\\scriptstyle
- 2 b \langle \{x, p\} \rangle^2
\end{array} \\[20pt]

\begin{array}{c}\scriptstyle0\end{array} &
\begin{array}{c}\scriptstyle0\end{array} &
\begin{array}{c}\scriptstyle
4 c \langle x^2 \rangle \langle \{x, p\} \rangle \\\scriptstyle
+ 2 \gamma \langle \{x, p\} \rangle^2
\end{array} &
\begin{array}{c}\scriptstyle
8 \gamma \langle p^2 \rangle^2 - \frac{8 \gamma T}{b} \langle p^2 \rangle \\\scriptstyle
+ 4 c \langle p^2 \rangle \langle \{x, p\} \rangle
\end{array} &
\begin{array}{c}\scriptstyle
8 c \langle x^2 \rangle \langle p^2 \rangle + 2 c + 2 c \langle \{x, p\} \rangle^2 \\\scriptstyle
+ 8 \gamma \langle p^2 \rangle \langle \{x, p\} \rangle - \frac{4 \gamma T}{b} \langle \{x, p\} \rangle
\end{array} \\[20pt]

\begin{array}{c}\scriptstyle0\end{array} &
\begin{array}{c}\scriptstyle0\end{array} &
\begin{array}{c}\scriptstyle
8 c \langle x^2 \rangle^2 \\
\scriptstyle + 2 b \\\scriptstyle
+ 4 \gamma \langle x^2 \rangle \langle \{x, p\} \rangle
\end{array} &
\begin{array}{c}\scriptstyle
-8 b \langle p^2 \rangle^2 - 2 c + 4 \gamma \langle p^2 \rangle \langle \{x, p\} \rangle \\\scriptstyle
+ 2 c \langle \{x, p\} \rangle^2 - \frac{4 \gamma T}{b} \langle \{x, p\} \rangle
\end{array} &
\begin{array}{c}\scriptstyle
8 \gamma \langle x^2 \rangle \langle p^2 \rangle - \frac{8 \gamma T}{b} \langle x^2 \rangle - 2 \gamma \\\scriptstyle
+ 8 c \langle x^2 \rangle \langle \{x, p\} \rangle  - 8 b \langle p^2 \rangle \langle \{x, p\} \rangle\\\scriptstyle
+ 2 \gamma \langle \{x, p\} \rangle^2
\end{array}
\end{bmatrix}
\label{eq:GSM_matrix_stacked}
\end{equation}
\end{widetext}
while the constant terms remain the same:
\begin{align}
D_T = \left[ 0,0,0, - 2\frac{\gamma}{b},0\right].
\end{align}
Since we have initialized the system with $\langle\hat x\rangle=\langle\hat p\rangle=0$ the factorization between first order and second order is still present in this squeezed Gaussian state. 

Let us now proceed to explicitly solve for the expansion coefficients of the temperature SLD by first computing the relevant Ehrenfest equations of motion \cite{breuer2002theory} associated with the CL master equation
\begin{align}
&\frac{d}{dt} \langle x^2 \rangle = \frac{1}{m} \langle px + xp \rangle, \\
&\frac{d}{dt} \langle px + xp \rangle = \frac{2}{m} \langle p^2 \rangle - 2 \langle  m\omega^2 x^2 \rangle - 2\gamma \langle px + xp \rangle, \\
&\frac{d}{dt} \langle p^2 \rangle = - m\omega^2\langle p x + x p \rangle - 4 \gamma \langle p^2 \rangle + 4 m \gamma T.
\end{align}
We start from the same initial conditions of $\langle x^2\rangle\langle p^2\rangle=1$, $\langle \{x,p\}\rangle=1$. We choose two sets of evolution parameters that correspond to two qualitatively different scenarios. Making sure that $\gamma \ll 2\pi T$ and $\omega\ll 2\pi T$ we consider a probe particle within a bath at temperature $T/m=4$ and relaxation rate $\gamma/m=1/8$. In one case the confining potential is given by $\omega/m=4/3$ in the other by $\omega/m=2/3$.

\begin{figure*}[t]
\centering
\includegraphics[scale=0.35]{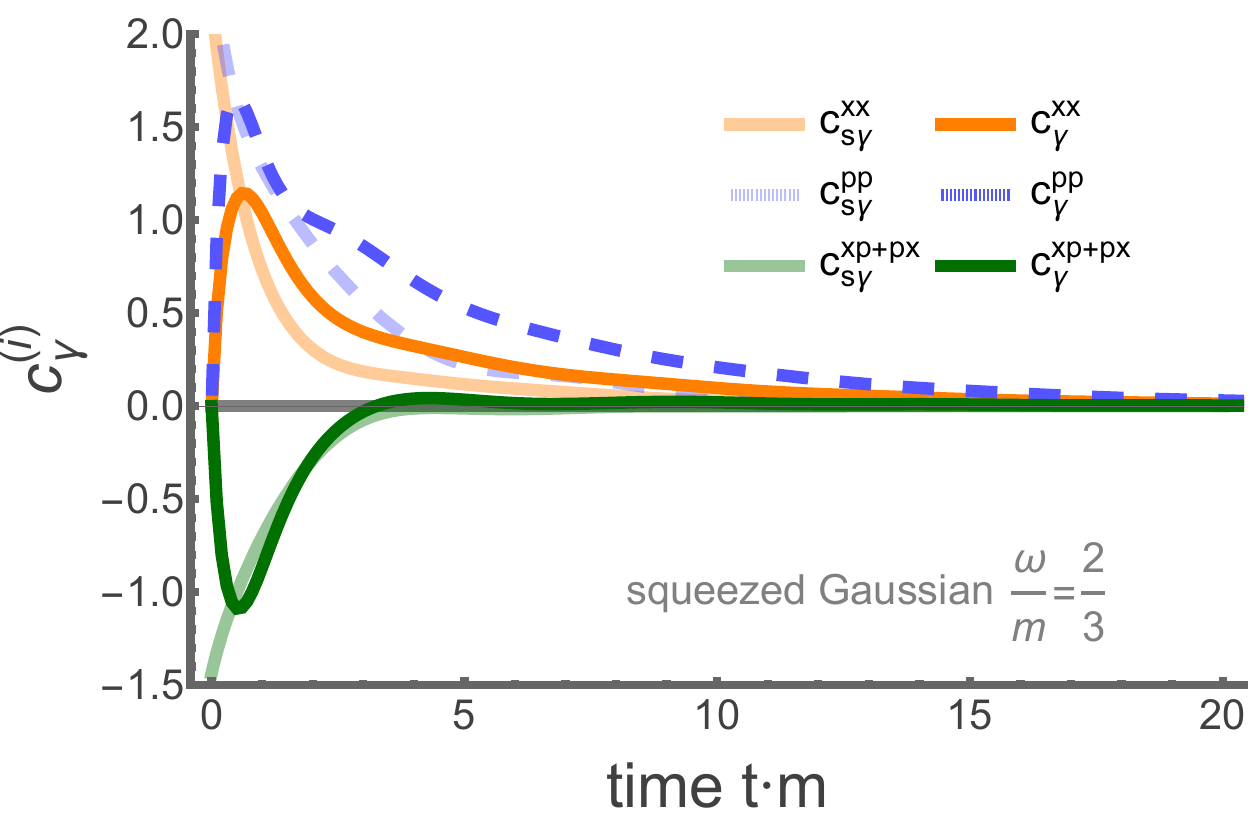}
\includegraphics[scale=0.35]{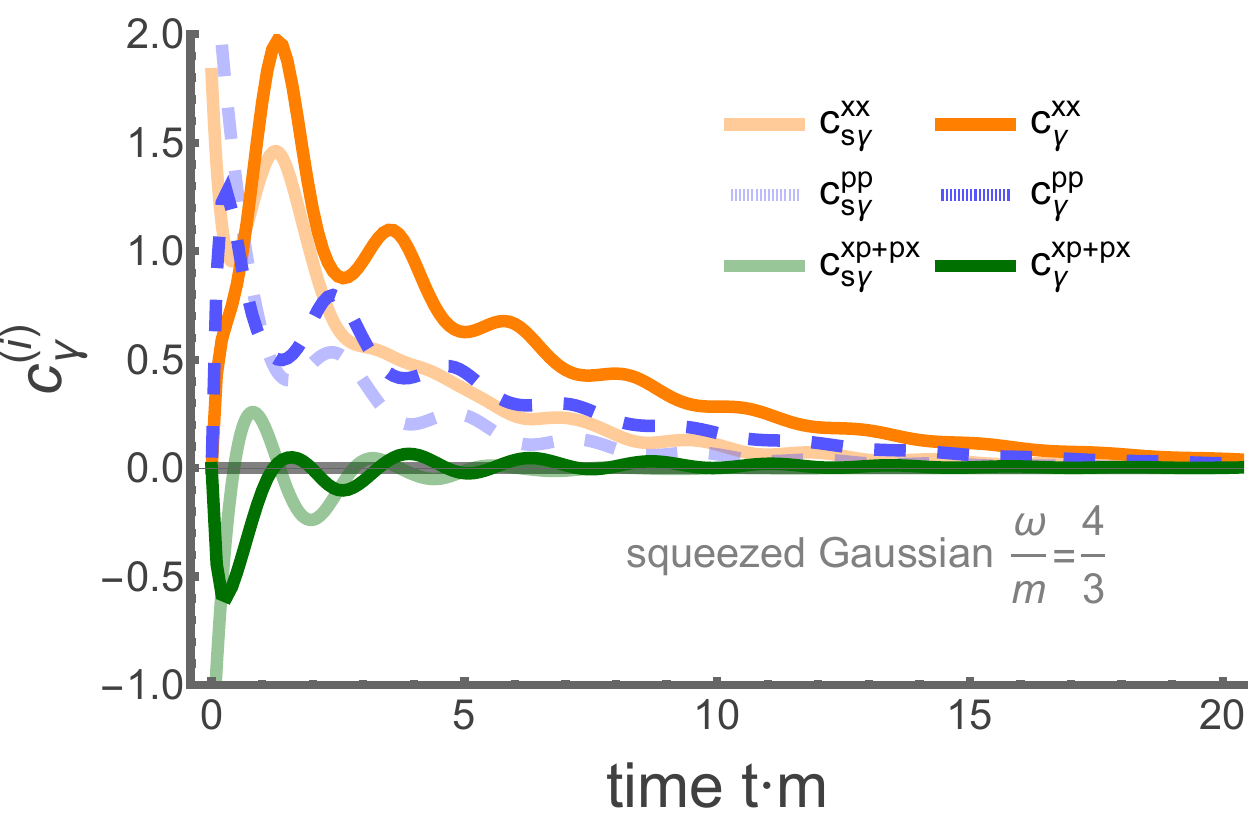}
\caption{Probe particle evolution in a squeezed Gaussian state at $\omega/m=2/3$ (left) and $\omega/m=4/3$ (right). Evolution of the expansion coefficients $c^{(i)}_\gamma$ of the symmetric logarithmic derivative for relaxation rate $c_\gamma^{xx}$ (orange solid), $c_\gamma^{pp}$ (blue dashed), $c_\gamma^{xp+px}$ (green solid). All coefficients vanish in the late time stationary state. Here we show the coefficients obtained from the solution of the defining differential equation \cref{eq:linsysc} as opaque lines with the coefficients obtained from evaluating simply the stationary approximation \cref{eq:linsysc2} at the same time step (transparent lines).} \label{fig:sqGaussEvolgamma}
\end{figure*}

In the potential with stronger curvature $\omega/m=4/3$, the particle is unable to diffuse over similarly large distances as in the weaker potential $\omega/m=2/3$. Thus we expect differences in position and momentum spread to affect the composition of the SLD.

Let us first test this intuition by inspecting the expectation values of the relevant operators in the left column of plots in \cref{fig:sqGaussEvol}. We provide the results for $\omega/m=2/3$ in the top row, those for $\omega/m=4/3$ in the bottom row. Indeed we find that for $\omega/m=4/3$ (bottom row) momentum spread (blue dashed) dominates over position spread (orange solid), in contrast to the less confining potential with $\omega/m=2/3$ (top row). Reassuringly at late times all expectation values relax to their equilibrium values given as gray solid lines. 

In addition to the evolution of the system expectation values, we now solve the defining differential equation \cref{eq:linsysc}. To this end we deploy a Magnus-type solver for matrix valued differential equations (for details see, e.g., Ref.~\cite{Puzzuoli:2022zbj}). An explicit implementation of Ehrenfest equations of the evolution equations for the SLD expansion coefficients can be found in the accompanying \texttt{Mathematica} scripts available open-access at the Zenodo repository \cite{rothkopfZ:2025}.

Since we assume the probe to be ignorant of the environment before coming into contact with it, the natural initial conditions are $c_\theta^{(i)}=0$. The results we obtain are shown in the right column of plots in \cref{fig:sqGaussEvol}. The coefficients obtained from the solution of the differential equation \cref{eq:linsysc} are given as opaque lines. By construction all of them vanish at inital time and converge to the correct equilibrium values at late times. For completeness we also show as transparent lines the coefficients obtained from the stationary approximation, i.e. from solving the algebraic system of equations \cref{eq:linsysc2}. Note that they consistently overestimate the magnitude of the true values at initial time, where it is the matrix $G$, not the matrix $M$, that governs $c^{(i)}_\theta$.

Note that the expansion coefficients $c_T^{(i)}$ of the SLD (right column) appear to take on values which weight the different operators roughly such that they contribute equally in the linear combination. That is, when momentum spread dominates position spread by roughly a factor of two (bottom row), the coefficient $c^{xx}_T$ is also roughly twice as large as $c^{pp}_T$. This reinforces the picture that it truly is the linear combination of position {\it and} momentum spread that constitutes the most sensitive probe of temperature. Similarly when position spread dominates, the coefficient $c^{pp}_T$ takes on a larger value than $c^{xx}_T$ (top row). 

As all expansion coefficients in \cref{fig:sqGaussEvol} approach the correct equilibrium values in the late time limit, we are confident that our results are meaningful in the transient regime too.

\subsection{Relaxation rate metrology}

\begin{figure*}[t]
\centering
\includegraphics[scale=0.35]{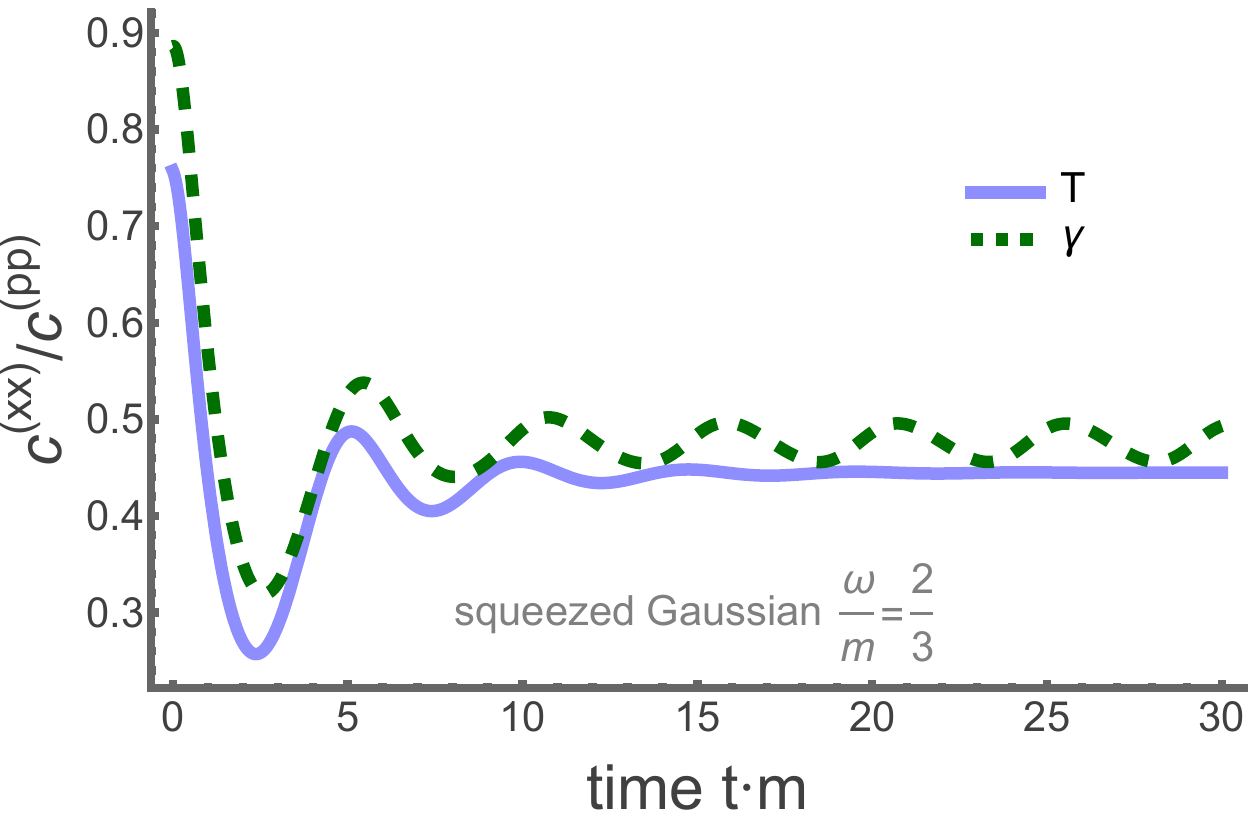}
\includegraphics[scale=0.35]{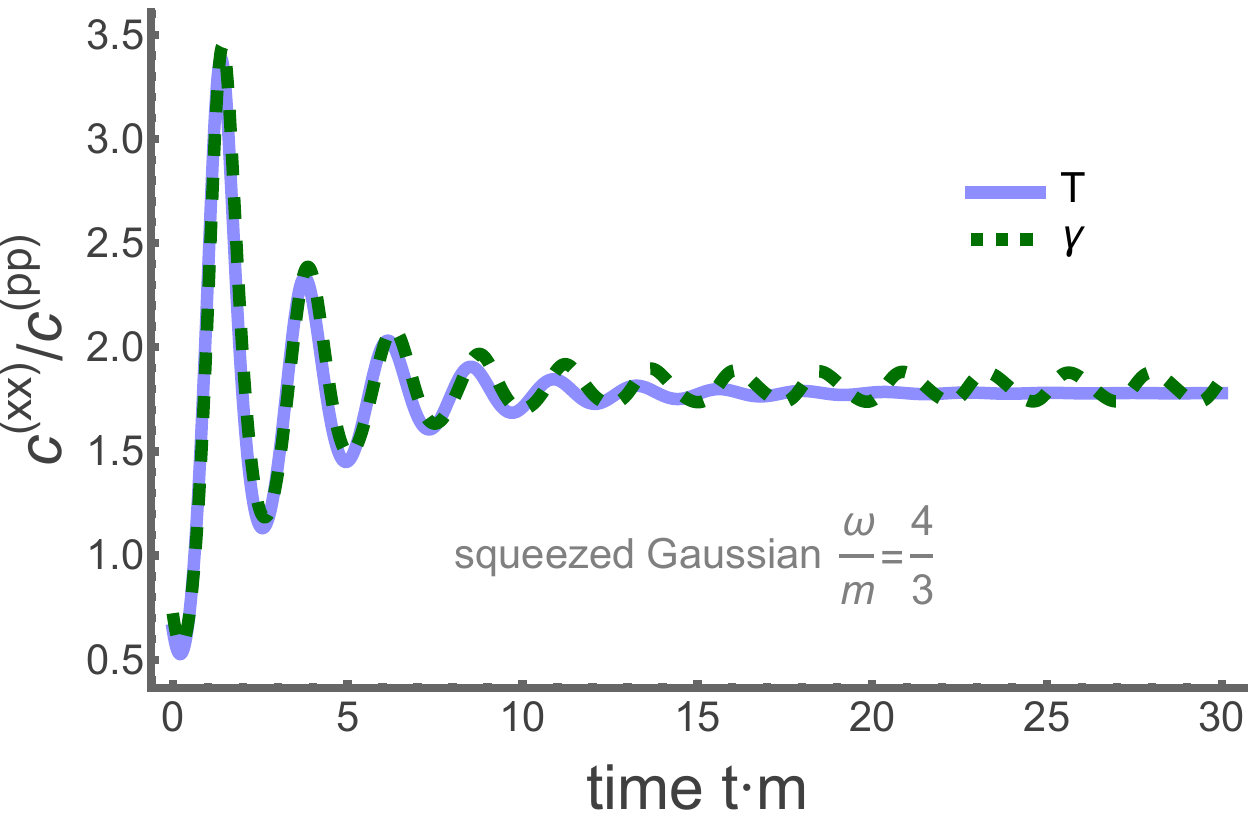}
\caption{Evolution of the ratio of expansion coefficients $c^{xx}/c^{pp}$ of the symmetric logarithmic derivative for temperature (green solid) and for the relaxation rate (purple solid) for the two cases of $\omega/m=2/3$ (left) and $\omega/m=4/3$ (right).} \label{fig:sqGaussEvolratios}
\end{figure*}

Having established that our formalism reproduces the correct temperature SLD in equilibrium and produces predictions that are consistent with physical intuition, let us deploy it for a genuine non-equilibrium estimate, i.e. for constructing the SLD for the relaxation rate of the probe.   

The CL master equation carries explicit factors of $\gamma$ in two terms, in contrast to temperature, which appears only once. But since the relaxation rate enters solely as prefactor to these terms, taking the derivative $\partial_\gamma(\partial_t \hat \rho)$ leads to the same terms in the system matrix $M$ as in \cref{eq:Mij} and the only difference lies in one additional term contributing to the constant contribution $D$
\begin{align}
    D_j=\frac{T}{b}\left\langle \left[ \hat x,\left[ \hat x, \hat A_j \right] \right]\right\rangle + i\left\langle \left\{ \hat p,\left[\hat A_j,\hat x\right]\right\}\right\rangle. \label{eq:Djgamma}
\end{align}

Considering the same squeezed Gaussian state as before, we have as system matrix \cref{eq:GSM_matrix_stacked} and for the constant terms we find
\begin{align}
D_\gamma = \left[ 0,0,0, - 2\frac{T}{b}+4 \left\langle p^2 \right\rangle, 2\left\langle \{\hat x,\hat p\} \right\rangle \right].
\end{align}

The evolution of the coefficients is shown in \cref{fig:sqGaussEvolgamma} for $\omega/m=2/3$ (left) and $\omega/m=4/3$ (right). As expected from the fact that in the steady state at late time no information about equilibration is retained, all coefficients approach zero over time. Again the ordering of the coefficients $c_\gamma^{xx}$ and $c_\gamma^{pp}$ follows the inverted hierarchy of the expectation values.

For completeness we here also include as transparent lines the coefficients obtained from the stationary approximation, i.e. from solving the algebraic system of equations \cref{eq:linsysc2}. As before they overestimate the correct values at early times.

\begin{figure*}[t]
\centering
\includegraphics[scale=0.35]{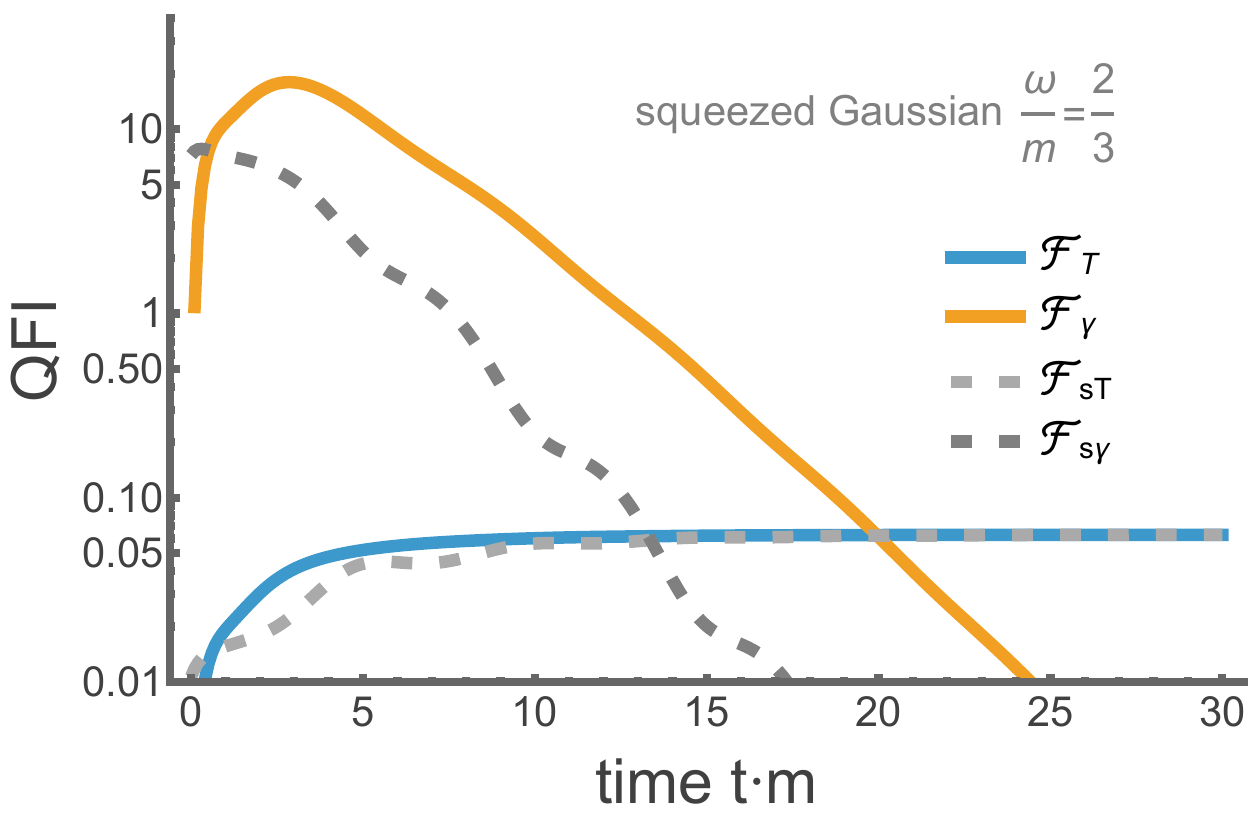}
\includegraphics[scale=0.35]{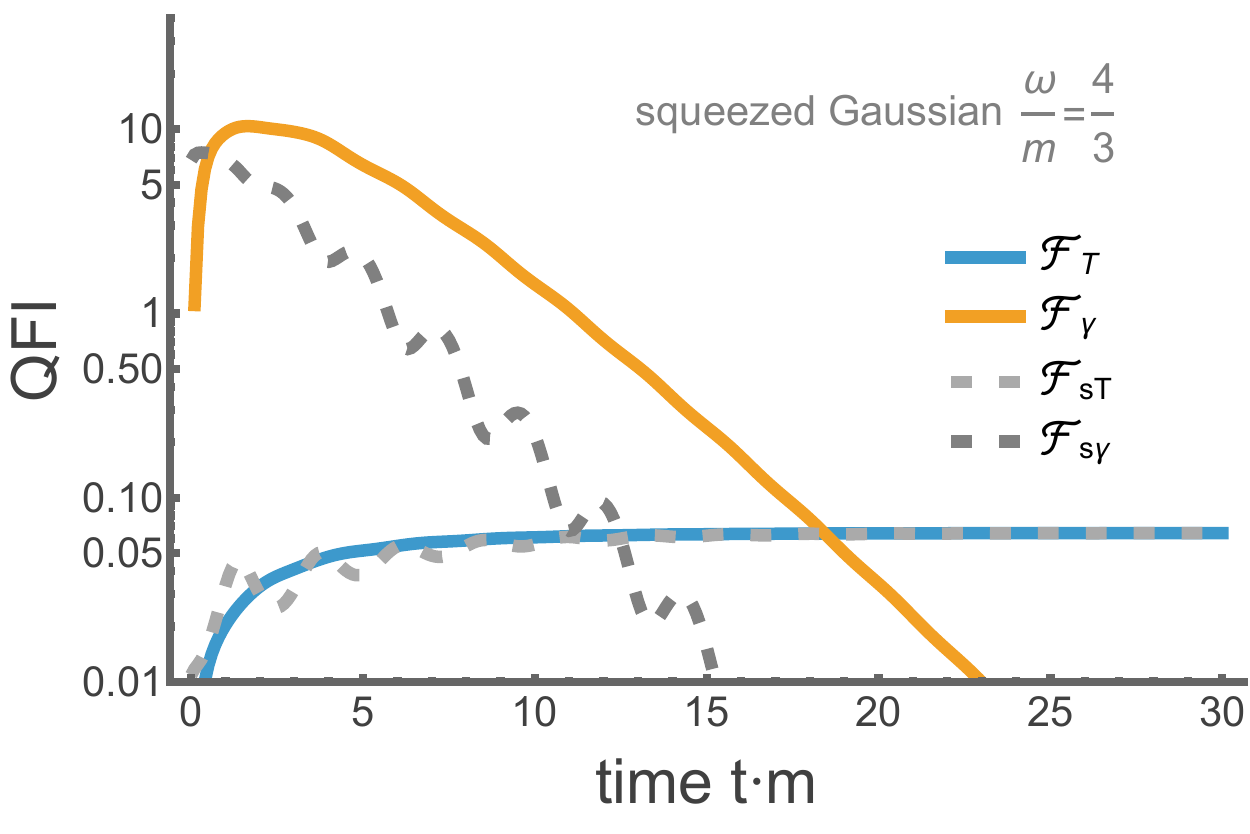}
\caption{Evolution of the quantum Fisher information for temperature measurements (blue solid) and for relaxation rate measurements (orange solid) for the two cases of $\omega/m=2/3$ (left) and $\omega/m=4/3$ (right). Note the qualitatively different behavior with ${\cal F}_T$ increasing towards late times, while ${\cal F}_\gamma$ drops to zero. As gray dashed lines we also show the resulting estimate of the QFI arising from the stationary approximation.} \label{fig:sqGaussEvolQFI}
\end{figure*}

Due to the simple dependence of the CL master equation on the environmental parameters $\gamma$ and $T$ we find that a similarly weighted linear combination of position and momentum spread constitutes an optimal estimator for both of these properties. Of course out of equilibrium, the mixed operator $\{x,p\}$ does contribute to the SLD but its expectation value decays towards zero. \Cref{fig:sqGaussEvolratios} shows the explicit ratios of the coefficients $c^{xx}/c^{pp}$ for $\hat \Lambda_T$ (blue solid) and $\hat \Lambda_\gamma$ (green dashed) evaluated for $\omega/m=2/3$ (left) and $\omega/m=4/3$ (right). We find that for the parameters considered here, the ratios of these coefficients lie close together throughout the evolution towards the steady state.

\begin{figure*}[t]
\centering
\includegraphics[scale=0.35]{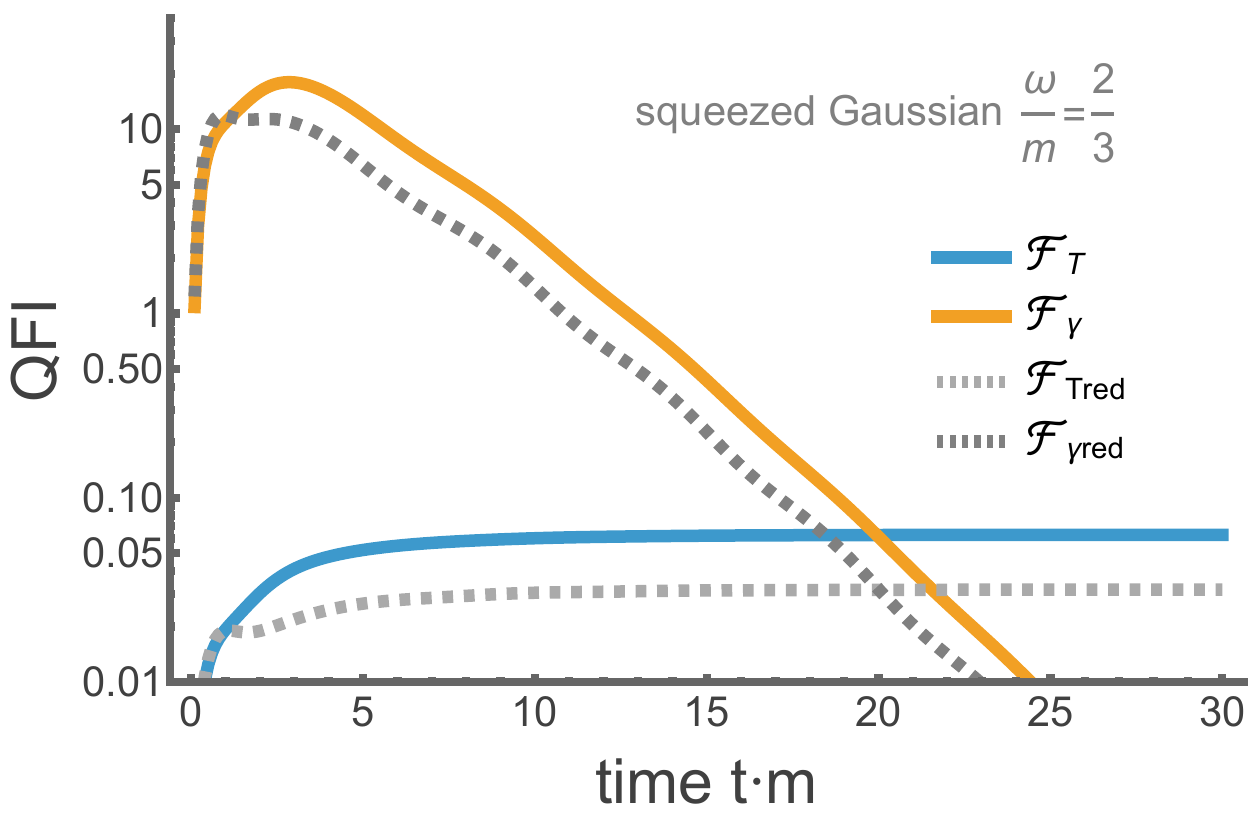}
\includegraphics[scale=0.35]{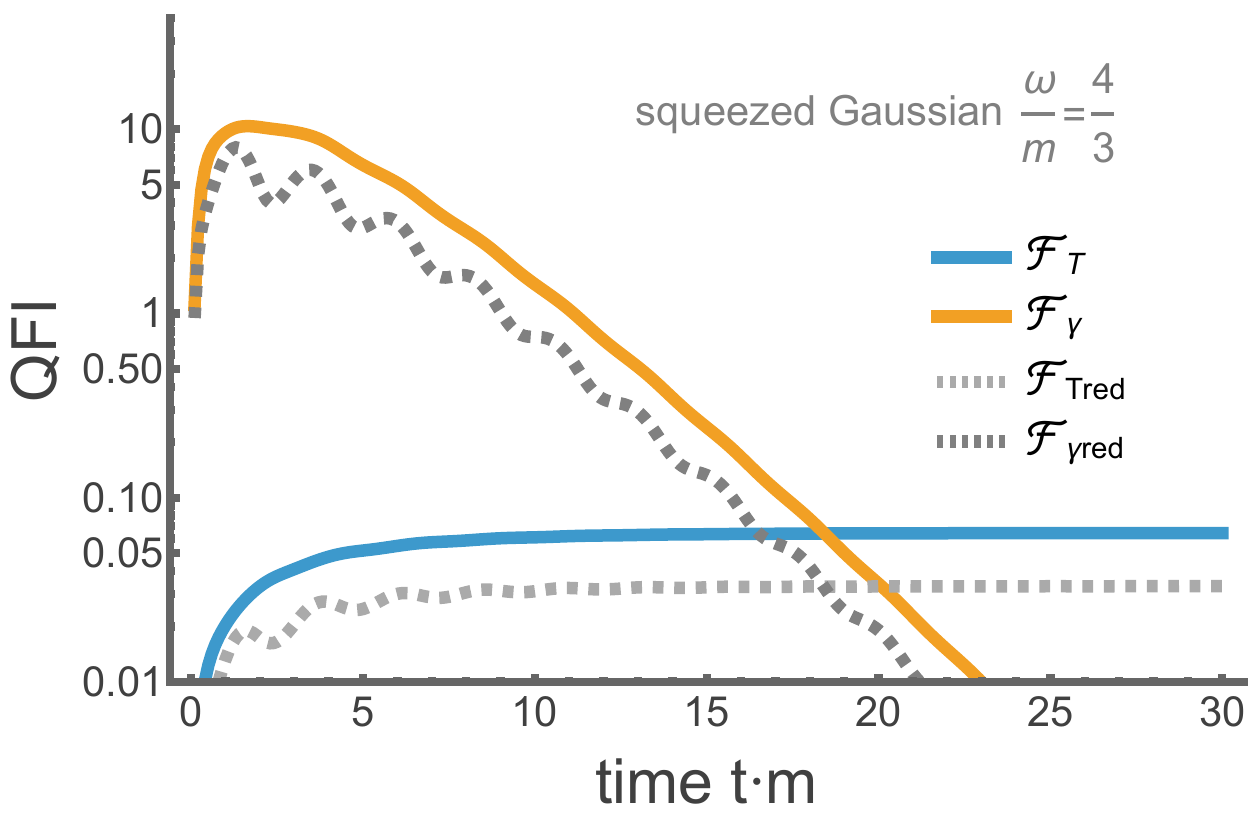}
\caption{Dependence on the available basis of observables of the achievable quantum Fisher information for temperature measurements (blue solid) and for relaxation rate measurement (orange solid) for the two cases of $\omega/m=2/3$ (left) and $\omega/m=4/3$ (right). We find that removing the $\hat p^2$ operator from the basis by setting its expansion coefficient to zero (gray dashed lines) monotonously reduces the achievable QFI.} \label{fig:sqGaussEvolQFIbasis}
\end{figure*}

{\subsection{Quantum Fisher Information}}

The explicit construction of the SLD is only part of the necessary steps to perform optimal temperature and relaxation rate metrology, as we also need to understand in which time intervals measurements are feasible. Intuitively, we expect that a fully thermally equilibrated probe will be most informative about the environment temperature, while it has lost all knowledge of the approach to equilibrium and thus becomes less and less informative about the relaxation rate.

{This intuition can be made precise by constructing the explicit expression for the quantum Fisher information, defined as the variance of the SLD
\[
{\rm QFI} \equiv{\cal F} = \mathrm{Var}(\Lambda)
=
\langle \Lambda^2 \rangle - \langle \Lambda \rangle^2 = \sum_{i,j} c^{(i)} G_{ij} c^{(j)}.
\]

And indeed, inspecting the quantum Fisher information ${\cal F}$ for temperature ${\cal F}_T$ and relaxation rate ${\cal F}_\gamma$, as shown in \cref{fig:sqGaussEvolQFI} confirms our intuition. The QFI for temperature steadily increases until it has reached its maximum value at late times. On the other hand for the relaxation rate measurement we find two regimes. At first the probe must establish contact with the environment during which the sensitivity to $\gamma$ rapidly increases. Thereafter, however, the relaxation-rate QFI decreases monotonically over time, rendering the SLD for $\gamma$ irrelevant in the late time limit. 

When comparing to the result from the static approximation, given as gray dashed lines, we find that as expected it initially overestimates the amount of information we can learn about the environment but subsequently fails to reach a sensitivity comparable to the true solution derived from the differential equation containing both $G$ and $M$.

As a final step, let us explore the dependence of the achievable quantum Fisher information on the available basis operators. From the original definition of the SLD we are guaranteed that the operator $\Lambda$ will extract the maximum possible amount of information about the environmental parameter $\theta$ permitted by the Cram\'er-Rao bound. This however is only achieved if a complete basis of operators $\hat A_i$ is provided. 

In practice, experimental constraints limit the basis of accessible observables and we can meaningfully ask only for the maximum information achievable from those accessible observables. In \cref{fig:sqGaussEvolQFIbasis} we show the result for the QFI using the set of observables $\hat x^2$, $\hat p^2$ and $\{\hat x,\hat p\}$ as solid lines. Removing the $\hat p^2$ operator from the basis by setting its expansion coefficient to zero clearly reduces the value of the achievable QFI, shown as gray dotted lines. At the same time, by construction, we are guaranteed that by adding more and more operators, completing the basis $\hat A_i$, we recover an SLD that indeed saturates the quantum Cram\'er-Rao bound.\\

\section{Summary and Conclusion}
\label{sec:sum}

We have introduced a general strategy for the explicit construction of optimally sensitive observables for quantum probes, based on the symmetric logarithmic derivative. By exploiting the symmetry of mixed derivatives guaranteed by Schwarz's theorem, we relate the dynamical evolution governed by a master equation to the parameter sensitivity encoded in the SLD. Employing an operator basis expansion, our approach translates the identification of the SLD into the solution of a system of linear ordinary differential equations for the associated expansion coefficients. In equilibrium the determination reduces to the solution of a linear system of algebraic equations for the expansion coefficients.

To validate the method, we applied it to the Caldeira–Leggett model in the high-temperature limit, where analytic expressions for the steady-state density matrix are available. In this setting, our construction successfully recovers the known SLD associated with temperature estimation in the late-time Gaussian state.

Crucially, our framework enables the determination of the SLD directly from the master equation, without requiring an explicit solution for the time-evolved density matrix. All relevant information is encoded in the expectation values of experimentally accessible observables. This novelty allows us to construct time-dependent expressions for the SLD even when the probe is out of equilibrium. We find that the structure of the optimal observable evolves over time and asymptotically approaches its equilibrium form in the long-time limit.

Access to the SLD in the transient regime opens new directions for quantum metrology beyond equilibrium. As an illustrative example, we derive the SLD associated with the environmental relaxation rate. Owing to the simple dependence of the Caldeira–Leggett dynamics on both relaxation rate and temperature, this observable turns out to be dominated by a linear combination of the position and momentum variances, which mirrors the structure of the temperature SLD.

With explicit SLD expressions at hand, we compute the corresponding quantum Fisher information, revealing that the thermalized probe indeed exhibits maximal sensitivity to temperature, whereas sensitivity to the relaxation rate is concentrated in the non-equilibrium regime. These findings highlight the potential of our approach to extend the practical applicability of the SLD in quantum metrology, particularly in regimes where the reduced state of the probe is not analytically accessible.

\acknowledgments
We thank Alfonso V. Ramallo and Mahn-Soo Choi for insightful discussions.
A.R.~thanks Korea University for support through project K2605081 {\it Ab-initio lattice simulations of the real-time dynamics of non-relativistic fermions,} as well as projects K2503291, K2511131 and K2510461. VLP was supported by Xunta de Galicia project ED481A 2022/286, by European Research Council project ERC-2018-ADG-835105 YoctoLHC, by Xunta de Galicia (CIGUS Network of Research Centres), by European Union ERDF, by the Spanish Research State Agency under projects PID2020-119632GBI00 and PID2023-152762NB-I00, and project CEX2023-001318-M financed by MCIN/AEI/10.13039/501100011033

\bibliographystyle{apsrev4-2}
\bibliography{OptimalObservables}

\begin{widetext}
    
\appendix

\section{Sketch of the derivation of the system matrix M}
\begin{table*}[htp!]
\begin{tabular}{|c|c|c|c|c|c|c|}
\hline
                  &  $ \scriptstyle \hat{x}$   & $\scriptstyle\hat{p}$                           & $\scriptstyle\hat{x}^2$          & $\scriptstyle\hat{x}\hat{p}$                          & $\scriptstyle\hat{p}^2$          & $\scriptstyle\hat{p}\hat{x}$ \\ \hline
$\scriptstyle\hat{x}^2$       &  $\scriptstyle\hat{x}^3$ & $\scriptstyle W(\hat{x}^2\hat{p})+i\hat{x}$      & $\scriptstyle\hat{x}^4$          & $\scriptstyle W(\hat{x}^3\hat{p})+\frac{3i}{2}\hat{x}$ & $\scriptstyle W(\hat{x}^2\hat{p}^2)+2iW(\hat{x}\hat{p})-\frac{1}{2}$& $\scriptstyle W(\hat{x}^3\hat{p})+\frac{i}{2}\hat{x}$\\ \hline
$\scriptstyle \hat{x}\hat{p}$  & $\scriptstyle W(\hat{x}^2\hat{p})$ & $\scriptstyle W(\hat{x}\hat{p}^2)+i\hat{p}$  &       $\scriptstyle W(\hat{x}^3\hat{p})-\frac{i}{2}\hat{x}^2$               &  $\scriptstyle W(\hat{x}^2\hat{p}^2)+iW(\hat{x}\hat{p})$   &    $\scriptstyle W(\hat{x}\hat{p}^3)+\frac{3i}{2}\hat{p}^{2}$   &     $\scriptstyle W(\hat{x}^2\hat{p}^2)+\frac{1}{2}$        \\ \hline
$\scriptstyle \hat{p}^2$       & $\scriptstyle W(\hat{x}\hat{p}^2)-i\hat{p}$  &   $\scriptstyle \hat{p}^3$           &     $\scriptstyle W(\hat{x}^2\hat{p}^2)-2iW(\hat{x}\hat{p})-\frac{1}{2}$  &     $\scriptstyle W(\hat{x}\hat{p}^3)-\frac{i}{2}\hat{p}^{2}$             &     $\scriptstyle \hat{p}^{4}$                 &    $\scriptstyle W(\hat{x}\hat{p}^3)-\frac{3i}{2}\hat{p}^{2}$          \\ \hline
$\scriptstyle \hat{p}\hat{x}$  &      $\scriptstyle W(\hat{x}^2\hat{p})-i\hat{x}$        &    $\scriptstyle W(\hat{x}\hat{p}^2)$  &    $\scriptstyle W(\hat{x}^3\hat{p})-\frac{3i}{2}\hat{x}^2$  & $\scriptstyle W(\hat{x}^2\hat{p}^2)+\frac{1}{2}$ &   $\scriptstyle W(\hat{x}\hat{p}^3)+\frac{i}{2}\hat{p}^{2}$ & $\scriptstyle W(\hat{x}^2\hat{p}^2)-iW(\hat{x}\hat{p})$ \\ \hline
\end{tabular}
\caption{Weyl correspondences for every combination of three or four operators appearing in the derivation of the matrix elements $M_{ij}$. The operator on the left column times the operator on the top row is equal to the corresponding cell, e.g.: $\hat{x}\hat{p}^2\hat{x}=(\hat{x}\hat{p})(\hat{p}\hat{x})=W(\hat{x}^2\hat{p}^2)+\frac{1}{2}$}\label{tab:Weyl}
\end{table*}

To obtain the expression \cref{eq:Mij} of the matrix elements $M_{ij}$ in terms of expectation values, we start by taking the temperature derivative of the operator-valued CL master equation \cref{eq:CLmaster}
\begin{align}
    \partial_T\partial_t \hat\rho(t,T) &= -i [\hat H,\partial_T\hat\rho] - i\gamma [\hat x,\{\hat p,\partial_T\hat\rho\}]\label{eq:DTCLmaster}-2\gamma m[ \hat x,[\hat x,\hat \rho]]-2\gamma m T[ \hat x,[\hat x,\partial_T\hat \rho]],
\end{align}
as well as the time derivative of the operator definition \cref{eq:SLDdef} of the SLD, setting $\theta=T$, which yields
\begin{align}
\partial_t\partial_T \hat \rho(T)&= \frac{1}{2} \left( \hat\Lambda_T \partial_t\hat\rho(T) + \partial_t\hat\rho(T) \hat\Lambda_T \right)\label{eq:DtSLDdef}- \partial_t\hat\rho(T)\langle \hat \Lambda_T \rangle- \hat\rho(T)\partial_t\langle \hat \Lambda_T \rangle,
\end{align}
where $\partial_t\langle \hat \Lambda_T \rangle=\operatorname{Tr}[\partial_t\hat\rho(T)\hat \Lambda_T]$. Now, due to Schwarz's theorem, we have \cref{eq:identify} and we can identify \cref{eq:DTCLmaster} and \cref{eq:DtSLDdef}. 

Since we are interested here in the derivation of the system matrix $M$ only, we insert \cref{eq:CLmaster} and \cref{eq:SLDdef} and discard any time derivatives. To convert to expectation values, we make the operator expansion ansatz $\hat \Lambda_T = \sum_i c^{(i)}_T \hat A_i$ as in \cref{eq:ansatzsld} and take traces according to \cref{eq:convexpval}. This allows us to obtain \cref{eq:linsysc2} with $M_{ij}$ and $D_j$ as given by \cref{eq:Mij} and \cref{eq:Dj}, respectively. 

We can evaluate the individual entries of $M_{ij}$, for example $M_{x^2p^2}$ by inserting the corresponding expressions for $\hat A_i$ and $\hat A_j$ in \cref{eq:Mij}
\begin{align}
&M_{x^2p^2} =\label{eq:Mx2p2}\\ 
& -\tfrac{1}{2} \left\langle \left\{ \hat{x}^2, \left[ \hat{H}, \hat{p}^2 \right] \right\}\right\rangle 
- \tfrac{i}{2} \left\langle \left\{ \hat{x}^2, \left\{ \hat{p} ,\left[ \hat{p}^2, \hat{x} \right] \right\} \right\} \right\rangle \nonumber - \tfrac{\gamma T}{2 b} \left\langle \left\{ \hat{x}^2, \left[ \hat{x}, \left[ \hat{x}, \hat{p}^2 \right] \right] \right\} \right\rangle 
+ \tfrac{i \gamma}{2} \left\langle \left\{ \hat{p}, \left[ \left\{ \hat{x}^2, \hat{p}^2\right\}, \hat{x} \right] \right\}\right\rangle \nonumber\\
& + \tfrac{\gamma T}{2 b} \left\langle \left[ \hat{x}, \left[ \hat x, \left\{ \hat{x}^2, \hat{p}^2 \right\}\right] \right] \right\rangle 
- i \left \langle \hat{p}^2 \right\rangle \left\langle \left[ \hat{x}^2, \hat{H} \right] \right\rangle \nonumber- i \gamma \left\langle \hat{p}^2 \right\rangle \left\langle \left\{ \hat{p}, \left[ \hat{x}^2, \hat{x} \right] \right\} \right\rangle 
- \tfrac{\gamma T}{b} \left\langle \hat{p}^2 \right\rangle \left\langle \left[ \hat x, \left[ \hat{x}, \hat{x}^2 \right] \right] \right\rangle.
\end{align}
Here we make use of known commutation relations
\begin{align}
\nonumber &[\hat x,\hat p^m]= i m\hat p^{m-1}, \quad [\hat x^n,\hat p]= i n\hat x^{n-1}, \quad [\hat x, \{\hat x,\hat p\}]=2i\hat x, \quad [\hat p, \{\hat x,\hat p\}]=-2i\hat p, \quad [\hat x^2,\hat p^2]=2i\{\hat x,\hat p\},
\end{align} as well as the Weyl correspondences collected in \cref{tab:Weyl} to obtain
\begin{equation}
    M_{x^2p^2}=-4b\langle W(xp^3)\rangle+2\gamma+4b\langle p^2\rangle\langle W(xp)\rangle.
\end{equation}
This expression reproduces both the steady state result ($M_{x^2p^2}=2\gamma$) when using \cref{eq:Wxp} and \cref{eq:Wx3p}, as well as the squeezed Gaussian state result ($M_{x^2p^2}=-4b\langle p^2\rangle \langle\{\hat x, \hat p\}\rangle+2\gamma$) when using \cref{eq:Wx3pSqueezed}.

\section{Expansion coefficients of the temperature SLD in the squeezed Gaussian state}\label{sec:coeffs}

For completeness and reproducibility, we provide below the explicit expressions for the non-vanishing expansion coefficients of the temperature SLD in the squeezed Gaussian state from solving the set of linear algebraic equations \cref{eq:linsysc2}.

\begin{align}
c^{(x^2)}_T&=f(\langle x^2\rangle,\langle p^2\rangle,\langle \{x,p\}\rangle)=\frac{f_N(\langle x^2\rangle,\langle p^2\rangle,\langle \{x,p\}\rangle)}{\Delta(\langle x^2\rangle,\langle p^2\rangle,\langle \{x,p\}\rangle)}\\
c^{(p^2)}_T&=g(\langle x^2\rangle,\langle p^2\rangle,\langle \{x,p\}\rangle)=\frac{g_N(\langle x^2\rangle,\langle p^2\rangle,\langle \{x,p\}\rangle)}{\Delta(\langle x^2\rangle,\langle p^2\rangle,\langle \{x,p\}\rangle)}\\
c^{(\{x,p\})}_T&=h(\langle x^2\rangle,\langle p^2\rangle,\langle \{x,p\}\rangle)=\frac{h_N(\langle x^2\rangle,\langle p^2\rangle,\langle \{x,p\}\rangle)}{\Delta(\langle x^2\rangle,\langle p^2\rangle,\langle \{x,p\}\rangle)}
\end{align}
\begin{align}
    f_N&=18 \left[\gamma  \left\langle \left\{x,p\right\}\right\rangle  \left(-4 \left(9
   \left\langle p^2\right\rangle ^2-1\right) \left\langle x^2\right\rangle  \left(2
   T-\left\langle p^2\right\rangle \right)+9 \left\langle p^2\right\rangle ^2-18
   T^2+1\right)\right.\\&\left.
   +\gamma  \left\langle \left\{x,p\right\}\right\rangle {}^3 \left(18 T
   \left\langle p^2\right\rangle -9 \left\langle p^2\right\rangle ^2-18
   T^2+1\right)\right.\nonumber\\&\left.
   +\left(T-\left\langle p^2\right\rangle \right) \left(-9 \gamma ^2-4
   \left\langle x^2\right\rangle  \left(\left(1-9 \gamma ^2\right) \left\langle
   p^2\right\rangle +9 \left\langle p^2\right\rangle ^3+18 \gamma ^2 T\right)-9
   \left\langle p^2\right\rangle ^2-1\right)\right.\nonumber\\&\left.
   +\left\langle \left\{x,p\right\}\right\rangle {}^2
   \left(T-\left\langle p^2\right\rangle \right) \left(-9 \gamma ^2-4 \left\langle
   p^2\right\rangle  \left\langle x^2\right\rangle +9 \left\langle p^2\right\rangle
   ^2\right)+\left\langle \left\{x,p\right\}\right\rangle {}^4 \left(T-\left\langle
   p^2\right\rangle \right)\right]\nonumber\\
    g_N&=\tfrac{9}{2} \left[-9 \gamma  \left\langle \left\{x,p\right\}\right\rangle {}^3 \left(\left\langle
   x^2\right\rangle  \left(8 T-4 \left\langle p^2\right\rangle \right)+1\right)\right.\\\nonumber&\left.-72 \gamma
    \left\langle x^2\right\rangle  \left\langle \left\{x,p\right\}\right\rangle 
   \left(T-\left\langle p^2\right\rangle \right) \left(4 T \left\langle x^2\right\rangle
   +1\right)\right.\\&\left.-\left\langle \left\{x,p\right\}\right\rangle {}^2 \left(T-\left\langle
   p^2\right\rangle \right) \left(8 \left\langle x^2\right\rangle  \left(9 \left\langle
   p^2\right\rangle -2 \left\langle x^2\right\rangle \right)+9\right)\right.\nonumber\\&\left.-\left(16
   \left\langle x^2\right\rangle ^2+9\right) \left(T-\left\langle p^2\right\rangle
   \right) \left(4 \left\langle p^2\right\rangle  \left\langle x^2\right\rangle
   +1\right)-9 \gamma  \left\langle \left\{x,p\right\}\right\rangle {}^5\right]\nonumber\\
    h_N&=9 \left[4 \gamma  \left\langle x^2\right\rangle  \left\langle
   \left\{x,p\right\}\right\rangle {}^2 \left(-9 \left\langle p^2\right\rangle ^2+18
   T^2-1\right)-\gamma  \left(16 \left\langle x^2\right\rangle ^2+9\right)
   \left(T-\left\langle p^2\right\rangle \right)\right.\\&\left.
   +\left\langle \left\{x,p\right\}\right\rangle
   {}^3 \left(4 \left\langle x^2\right\rangle  \left(\left\langle p^2\right\rangle
   -T\right)-9 \gamma ^2\right)
   +\left\langle \left\{x,p\right\}\right\rangle 
   \left(T-\left\langle p^2\right\rangle \right) \left(\left\langle p^2\right\rangle 
   \left(4 \left\langle x^2\right\rangle  \left(9 \left\langle p^2\right\rangle +4
   \left\langle x^2\right\rangle \right)\right.\right.\right.\nonumber\\&\left.\left.\left.+9\right)+4 \left(1-9 \gamma ^2\right)
   \left\langle x^2\right\rangle \right)+9 \gamma  \left\langle p^2\right\rangle 
   \left\langle \left\{x,p\right\}\right\rangle {}^4\right]\nonumber\\
   \Delta&=\gamma  \left[18 \gamma  \left\langle
   \left\{x,p\right\}\right\rangle {}^3 \left(\left\langle x^2\right\rangle  \left(-72 T
   \left\langle p^2\right\rangle +72 T^2-4\right)+9 T\right)\right.\\
   &\left.-18 \gamma  \left\langle
   \left\{x,p\right\}\right\rangle  \left(\left\langle p^2\right\rangle  \left(4 \left\langle
   x^2\right\rangle  \left(-9 \left\langle p^2\right\rangle  \left(4 T \left\langle
   x^2\right\rangle +1\right)+\left(72 T^2-4\right) \left\langle x^2\right\rangle +18
   T\right)-9\right)\right.\right.\nonumber\\&\left.\left.+9 T-4 \left\langle x^2\right\rangle \right)-9 \left\langle
   \left\{x,p\right\}\right\rangle {}^4 \left(9 \left(\gamma ^2+\left\langle p^2\right\rangle
   ^2\right)+4 \left\langle x^2\right\rangle  \left(2 T-3 \left\langle p^2\right\rangle
   \right)\right)\nonumber+\left\langle \left\{x,p\right\}\right\rangle {}^2 \left(9 \left(9 \gamma
   ^2+2\right)\right.\right.\\&\left.\left.+4 \left\langle x^2\right\rangle  \left(9 \left\langle p^2\right\rangle 
   \left(9 \gamma ^2+4 \left\langle x^2\right\rangle  \left(4 T-3 \left\langle
   p^2\right\rangle \right)+9 \left\langle p^2\right\rangle  \left(\left\langle
   p^2\right\rangle -2 T\right)+1\right)+18 T-4 \left\langle x^2\right\rangle
   \right)\right)\right.\nonumber\\&\left.-\left(16 \left\langle x^2\right\rangle ^2+9\right) \left(18 T
   \left\langle p^2\right\rangle -9 \left\langle p^2\right\rangle ^2-1\right) \left(4
   \left\langle p^2\right\rangle  \left\langle x^2\right\rangle +1\right)+162 \gamma  T
   \left\langle \left\{x,p\right\}\right\rangle {}^5-9 \left\langle \left\{x,p\right\}\right\rangle
   {}^6\right]\nonumber
\end{align}
\end{widetext}

\end{document}